\documentclass[aps,prd,superscriptaddress,nofootinbib,eqsecnum,twocolumn]{revtex4-2}

\pdfoutput=1

\usepackage{subfigure}
\usepackage{amssymb}
\usepackage{amsmath}
\usepackage[svgnames]{xcolor}
\usepackage{mathtools,slashed}
\usepackage{physics}
\usepackage[compat=1.1.0]{tikz-feynhand}
\usepackage{tikz}
\usepackage[utf8]{inputenc}
\usepackage{url}
\usepackage[colorlinks,citecolor=DarkGreen,linkcolor=DarkRed,urlcolor=DarkBlue]{hyperref}
\usepackage{float}
\usepackage{newtxtext}
\usepackage{newtxmath}
\usepackage{hyperref} 
\usepackage[makeroom]{cancel}
\usepackage{ulem}

\begin{document}

\title{Magnetic enhancement and Hall conductivity of excitonic insulators in a Gross--Neveu type model}

\author{William R. Tavares}
\email{tavares.william@ce.uerj.br}
\affiliation{Departamento de
  F\'{\i}sica Te\'{o}rica, Universidade do Estado do Rio de Janeiro, 20550-013
  Rio de Janeiro, RJ, Brazil}
\affiliation{CFisUC, Department of Physics, University of Coimbra, 3004-516 Coimbra, Portugal}

\author{Rudnei O. Ramos} \email{rudnei@uerj.br} \affiliation{Departamento de
  F\'{\i}sica Te\'{o}rica, Universidade do Estado do Rio de Janeiro, 20550-013
  Rio de Janeiro, RJ, Brazil}

\author{Nei Lopes} \email{nei@cbpf.br}
\affiliation{Centro Brasileiro de Pesquisas F\'{\i}sicas, Rua Dr. Xavier Sigaud 150, Urca, 22290-180 Rio de Janeiro, RJ, Brazil}

\begin{abstract}

We study the effects of a perpendicular magnetic field on the excitonic insulator (EI) phase in 
the semiconductor regime using an extended planar four-Fermi model. Within the large-$N$ 
approximation, we determine the coupled scalar and excitonic condensates at finite temperature, 
chemical potential, and magnetic field. The field enhances the EI condensate and raises its critical 
temperature, providing an excitonic realization of magnetic catalysis, while the scalar condensate 
remains constant throughout the EI phase. By contrast, the critical chemical potential depends 
nonmonotonically on the field because of the successive occupation of Landau levels. The magnetic 
field also shifts the mean-field tricritical point and enlarges the first-order region of the 
temperature--chemical-potential phase diagram. We further analyze the Hall conductivity and find 
that increasing the field reduces the number of plateaus and modifies the threshold for a finite 
Hall response. {}For the parameters considered, the emergence of the EI condensate is accompanied 
by a characteristic change in the Hall conductivity, including a field-dependent change of slope 
near a continuous transition. These results show that the combined phase structure and Hall response 
can provide complementary signatures of excitonic ordering in planar fermionic systems.

\end{abstract}

\maketitle

\section{Introduction}

The excitonic insulator (EI) is an interaction-driven electronic phase originating from the spontaneous 
formation and condensation of bound electron-hole pairs, namely, excitons. The possibility of such a state 
was first discussed by Mott~\cite{Mott} and Knox~\cite{Knox} and was subsequently developed in 
Refs.~\cite{Des,Keldysh,Jerome,Halperin,Zittartz}. It can emerge in systems with either a small 
overlap between the valence and conduction bands or a sufficiently small semiconductor band gap. 
In the semimetal regime, the instability involves weakly bound and strongly overlapping electron-hole 
pairs and is therefore analogous, at the level of the pairing mechanism, to the Bardeen-Cooper-Schrieffer (BCS) 
regime of superconductivity~\cite{BCS}. On the semiconductor side, the excitons are more tightly bound, 
and the resulting state is closer to the Bose-Einstein condensation limit. In particular, when the exciton 
binding energy becomes comparable to or larger than the single-particle band gap, the normal semiconductor 
ground state can become unstable towards the formation of a coherent excitonic phase~\cite{Knox}. 
Although the resulting electron-hole pairs carry no net electric charge and therefore do not produce 
an ordinary charge supercurrent as in a superconductor, their condensation is associated with a 
nontrivial electronic order and, depending on the realization, can support neutral or counterflow coherent transport.

The theoretical description of the EI has traditionally been based on models containing conduction 
and valence bands coupled through an attractive electron-hole interaction~\cite{Des,Keldysh,Jerome,Halperin,Zittartz,Bronold,Comte}. 
At the same time, the experimental identification of an EI remains challenging because the 
opening or enhancement of an electronic gap by itself does not uniquely establish an excitonic origin. 
Signatures compatible with excitonic ordering have been reported in several candidate systems, 
including InAs/GaSb bilayers~\cite{Du,Eisenstein}, heterobilayers of two-dimensional materials~\cite{Gupta}, 
semiconductor double layers~\cite{Ma}, monolayer WTe$_2$~\cite{Jia}, and Ta$_2$NiSe$_5$~\cite{Lu,Volkov,Sugimoto,Kim,Salvo,Wakisaka,Wakisaka2,Windgatter}. 
In a number of these systems, modifications of the low-energy electronic spectrum below a 
characteristic temperature have been interpreted as evidence for the development of an additional 
interaction-driven gap~\cite{Du,Lu,Eisenstein,Volkov,Sugimoto,Kim,Wakisaka,Wakisaka2,Windgatter}. 
Studies include first-principles calculations for monolayer transition-metal 
dichalcogenides~\cite{Varsano} and monolayer WTe$_2$~\cite{Sun}, investigations of excitonic 
instabilities in one-dimensional systems~\cite{Liu}, semiconductor quantum wells~\cite{Littlewood}, 
and doped Dirac systems~\cite{Lizardo}, as well as studies of magnetic field effects in the candidate EI
 Ta$_2$NiSe$_5$~\cite{Mazza}. More recently, the EI and chiral phase structures of an extended 
Gross-Neveu (GN) model in $(1+1)-$ and $(2+1)-$dimensions were analyzed within the large-$N$ approximation 
in both the small-band-overlap semimetal and small-gap semiconductor regimes~\cite{lopes}, but in the 
absence of an external magnetic field.

A magnetic field provides a particularly useful probe of excitonic physics because it affects 
both the single-particle spectrum and the formation of electron-hole bound states. Early theoretical 
studies of excitonic insulators in magnetic fields~\cite{Elliott,fenton,Yafet} emphasized two 
important effects. First, a perpendicular magnetic field reorganizes the conduction and valence 
spectra into Landau levels, thereby replacing the zero-field band separation by an effective 
field-dependent energy scale involving the corresponding cyclotron energies. Second, magnetic 
confinement modifies the relative electron-hole motion and can enhance the exciton binding energy. 
Consequently, the magnetic field can change the balance between the semiconductor gap and the 
binding energy that controls the excitonic instability. This led already in early studies to 
the suggestion that high magnetic fields could provide a favorable route to access the EI phase 
without the need to tune the system through high external pressures~\cite{fenton}.

The influence of magnetic fields on excitonic systems has received renewed experimental attention. 
A recent example is magnetic field tuning between weak- and strong-coupling excitonic regimes in 
graphite~\cite{Zhu}. Magnetic field-dependent measurements are also particularly valuable because 
they provide information beyond the mere observation of an electronic gap. In particular, the Hall 
response can help distinguish an excitonic state from conventional insulating phases and has been 
proposed as an important diagnostic of the EI state~\cite{Jia}. More generally, a perpendicular 
magnetic field introduces Landau quantization and strongly reorganizes the density of low-energy 
fermionic states, giving rise to phenomena ranging from metal-insulator transitions in semiconductor 
systems~\cite{mit1,mit2,mit3} to the quantum Hall effect~\cite{hall1,hall2}, magnetic field-dependent superconducting 
transport~\cite{supercond1,supercond2,supercond3}, and the characteristic magnetic response of graphene and other planar 
Dirac materials~\cite{Graphene}. It is therefore natural to ask how these magnetic effects modify 
an interaction-driven excitonic condensate and whether the resulting phase transitions leave 
identifiable signatures in transport observables.

{}Four-fermion field theories in $(2+1)$-dimensions provide a useful effective framework to address 
this question. In particular, the Gross-Neveu model~\cite{gn} and its extensions have been extensively 
employed to describe dynamical fermion mass generation and symmetry-breaking phenomena in planar 
systems. Their magnetic field dependence has been especially well studied, including perpendicular 
orbital magnetic fields~\cite{Klimenko1,Klimenko2,Klimenko3,Gusynin1,Gusynin2,Vshivtsev:1995ug,Zhukovsky:2000yd,Gomes:2023vvu}, Zeeman 
effects~\cite{Caldas:2009zz,Ramos:2013aia,Klimenko:2013gua}, superconducting channels~\cite{Klimenko:2012qi,Klimenko:2012tk,Gomes:2022dmf}, 
corrections beyond the mean field~\cite{Kneur:2013cva,Mauldin:2026lcj}, and lattice formulations~\cite{Lenz:2023wvk,Lenz:2023gsq}. 
An important result emerging from these studies is magnetic catalysis~\cite{Shovkovy:2012zn,Miransky:2015ava}: 
Landau quantization enhances the tendency toward interaction-driven fermion condensation and can 
therefore strengthen dynamically generated gaps. Although this phenomenon is conventionally 
discussed for the chiral condensate, the same underlying question naturally arises for an excitonic 
interaction channel.

In the present context, the extended GN model considered in this work should be regarded as a 
low-energy effective description of the interaction channel responsible for electron-hole coherence 
rather than as a microscopic representation of the long-range Coulomb interaction. Its usefulness 
lies in providing a simple  framework in which the excitonic and chiral order parameters can be 
treated self-consistently and their response to temperature, density, and magnetic field can be 
followed analytically and numerically. The extension introduced in Ref.~\cite{lopes} contains, 
in addition to the standard GN interaction, the channel
$G_e(\overline{\psi}i\gamma_5\psi)^2$, whose corresponding condensate is used to describe the EI order. 
This construction allows us to investigate whether the magnetic-catalysis mechanism familiar from chiral 
symmetry breaking has an analog in the excitonic sector and how such an effect is reflected in the phase 
structure and transport properties of the model.

Motivated by these considerations, in this work we study the semiconductor regime of the EI phase 
in the extended $(2+1)$-dimensional GN model of Ref.~\cite{lopes} in the presence of a constant 
magnetic field perpendicular to the plane. Using a Hubbard-Stratonovich transformation and the 
large-$N$ approximation, we derive the finite-temperature effective potential including the Landau-level 
spectrum and obtain coupled gap equations for the chiral and excitonic condensates. We then determine how 
the EI phase evolves as a function of the magnetic field, temperature $T$, and chemical potential $\mu$. 
Our main objective is to establish how Landau quantization modifies the excitonic instability found at $B=0$ 
in Ref.~\cite{lopes}, with particular emphasis on the magnetic field dependence of the EI condensate, the 
critical temperature and chemical potential, and the location and nature of the corresponding phase transitions.

A central result of our analysis is that a perpendicular magnetic field enhances the EI condensate, providing 
an excitonic counterpart of the magnetic-catalysis phenomenon familiar from the chiral sector. 
This enhancement modifies the finite-$T$ and finite-$\mu$ phase structure and shifts the boundaries 
that separate the excitonic and nonexcitonic phases. We find, in particular, that the chiral condensate 
remains constant throughout the region in which the EI condensate is present; its subsequent magnetic field 
dependence after the emergence of the EI phase reduces to behavior already extensively studied for 
massive and massless GN-type theories~\cite{Ramos:2013aia,Kneur:2006ht,Barducci:1994cb,Klimenko1988}. We therefore 
concentrate primarily on the excitonic sector. In addition, we calculate the Hall conductivity and investigate 
its correlation with the emergence of the EI condensate. This provides a complementary observable through 
which the excitonic phase and its magnetic field-induced modifications may be characterized.

In addition to its direct application to planar fermionic systems, the GN model is also a useful toy 
model for dynamical symmetry breaking in relativistic theories and shares important structural 
features with Nambu-Jona-Lasinio-type descriptions~\cite{njl1,njl2}. Thus, some of the magnetic field 
effects studied here may also be of interest in other strongly interacting fermionic systems. 
However, our emphasis throughout this work is on the excitonic-insulator interpretation of the extended model.

The paper is organized as follows. In Sec.~\ref{GN_model_2+1_d}, we review the extended GN model in $(2+1)$-dimensions 
and introduce the chiral and EI order parameters, together with the large-$N$ thermodynamic potential in the absence 
of a magnetic field. In Sec.~\ref{GN_model_2+1_d_mag_field}, we incorporate a perpendicular magnetic field through 
Landau quantization and derive the corresponding thermodynamic potential and coupled gap equations at finite temperature 
and chemical potential. In Sec.~\ref{numerical_results}, we present the numerical results for the magnetic field 
dependence of the EI condensate, its phase structure, and the Hall conductivity. Finally, in Sec.~\ref{conclusions}, 
we summarize our main results and discuss their implications. Three appendices are included where some of the more technical details are given. 

Throughout the paper, we use natural units $\hbar=c=k_B=1$, and we absorb the effective Dirac velocity $v_F$ in 
the definition of the spatial coordinates. When comparing our results with experimentally relevant scales, 
we restore the appropriate physical units.

\section{Extended Gross--Neveu model for excitonic condensation}
\label{GN_model_2+1_d}

In this section, we introduce the extended Gross--Neveu (GN) model used as an effective field theory 
description of excitonic condensation. We first consider the system in the absence of an external magnetic 
field and discuss the physical roles of the two interaction channels. The effects of a constant magnetic 
field perpendicular to the plane will be incorporated in the following section.
The model is defined in $(2+1)$-dimensions by the Lagrangian density
\begin{align}
{\cal L}
={}&
\sum_{j=1}^{N}
\bar{\psi}_j
\left(
i\gamma^\mu\partial_\mu-m
\right)
\psi_j
+
\frac{g_c}{2N}
\left(
\sum_{j=1}^{N}\bar{\psi}_j\psi_j
\right)^2
\nonumber\\
&+
\frac{g_e}{2N}
\left(
\sum_{j=1}^{N}
\bar{\psi}_j i\gamma_5\psi_j
\right)^2 ,
\label{Lagr}
\end{align}
where $\psi_j$, with $j=1,\ldots,N$, denotes the $N$ fermion species,
$\bar{\psi}_j=\psi_j^\dagger\gamma^0$, and the spacetime index is
$\mu=0,1,2$. The couplings have been written in the standard
large-$N$ form, with $g_i=N G_i$, $i=c,e$, held fixed in the
$N\rightarrow\infty$ limit.

We employ the four-component reducible representation of the Dirac
algebra appropriate for planar fermions, following
Ref.~\cite{chiralsym}. In addition to the matrices $\gamma^\mu$,
$\mu=0,1,2$, this representation admits matrices that anticommute with
all three $\gamma^\mu$, including $\gamma_5$. It therefore allows one
to construct the two distinct flavor-singlet fermion bilinears
appearing in Eq.~(\ref{Lagr}).

The first term in Eq.~(\ref{Lagr}) describes the propagation of the
planar fermionic quasiparticles. The parameter $m$ introduces an
explicit single-particle gap and allows the model to describe the
semiconductor regime of interest here. More precisely, for the
corresponding free Dirac spectrum, the positive- and negative-energy
branches are separated by $2|m|$ at zero momentum. The parameter $m$
therefore fixes the bare band-gap scale before interaction-induced
corrections are included.

The interaction proportional to $g_c$ is the conventional scalar GN
channel. For $m=0$, it may generate a nonvanishing expectation value
$\langle\bar{\psi}\psi\rangle$ and dynamically break the discrete
chiral symmetry of the model. The resulting scalar condensate gives a
dynamical contribution to the fermionic gap. For $m\neq0$, the chiral
symmetry is explicitly broken, but we shall retain the standard
terminology chiral condensate'' for the scalar order parameter.

The interaction proportional to $g_e$ is the additional channel
introduced in Ref.~\cite{lopes} to describe excitonic ordering. In the
two-sector representation used there, the four-component fermion may
be decomposed into effective valence- and conduction-band sectors. The
matrix $\gamma_5$ is off diagonal in this sector space, so that the
bilinear $\bar{\psi}i\gamma_5\psi$
contains interband combinations of the form
$c^\dagger_{\rm c}c_{\rm v}+{\rm H.c.}$, up to conventions concerning
the relative phase of the two sectors. A nonvanishing expectation
value of this operator therefore produces an off-diagonal fermionic
self-energy and represents spontaneous interband coherence. In
condensed-matter language, it describes coherent particle-hole pairing
between the effective valence and conduction bands and will be
identified with the excitonic-insulator (EI) order parameter.

The model should thus be understood as a low-energy effective
description of the interaction channels relevant to the EI
instability. In particular, the local coupling $g_e$ is not intended
to explicitly reproduce the microscopic long-range electron-hole
Coulomb interaction. Rather, it parameterizes the effective interaction
in the interband-coherence channel. The advantage of this formulation
is that the scalar and excitonic condensates can be treated
self-consistently as functions of temperature, chemical potential,
and, in the following sections, an external magnetic field.

{}For $m=0$, the model is invariant under the discrete chiral
transformation
$\psi_j\rightarrow\gamma_5\psi_j,\;
\bar{\psi}_j\rightarrow-\bar{\psi}_j\gamma_5$, 
in which both fermion bilinears in Eq.~(\ref{Lagr}) change sign,
whereas their squares remain invariant. A nonzero scalar or excitonic
condensate may therefore break this discrete symmetry dynamically.
When $g_c=g_e$, the symmetry is enlarged to a continuous chiral
rotation that mixes the scalar and pseudoscalar bilinears, and the
model corresponds to the chiral GN model~~\cite{gn}, or Nambu--Jona-Lasinio-type
model~\cite{njl1,njl2,Flachi}. The explicit mass $m\neq0$ breaks the
chiral symmetry and leaves a finite bare gap, as appropriate for the
small-gap semiconductor regime considered in this work.

The relative strengths of $g_c$ and $g_e$ determine which condensation
channel is energetically favored. As discussed in
Ref.~\cite{lopes}, the conventional scalar phase is favored in the
region in which the scalar interaction dominates, whereas a
sufficiently strong excitonic coupling can generate a nonvanishing EI
condensate. In particular, the parameter region with $g_e>g_c$ is the
one relevant for the excitonic phase studied below.

To implement the large-$N$ expansion, it is convenient to linearize
the four-fermion interactions by introducing the auxiliary scalar
fields $\sigma$ and $\eta$. The Hubbard--Stratonovich form of
Eq.~(\ref{Lagr}) is
\begin{align}
{\cal L}[\bar{\psi},\psi,\sigma,\eta]
={}&
\sum_{j=1}^{N}
\bar{\psi}_j
\left(
i\slashed{\partial}
-m-\sigma-i\eta\gamma_5
\right)
\psi_j
\nonumber\\
&-
\frac{N}{2g_c}\sigma^2
-
\frac{N}{2g_e}\eta^2 .
\label{Lagrsigmaeta}
\end{align}
The equations of motion for the auxiliary fields give
\begin{align}
\sigma
&=
-\frac{g_c}{N}
\sum_{j=1}^{N}
\bar{\psi}_j\psi_j ,
\label{eqsigma}\\
\eta
&=
-\frac{g_e}{N}
\sum_{j=1}^{N}
\bar{\psi}_j i\gamma_5\psi_j .
\label{eqeta}
\end{align}
Substituting Eqs.~(\ref{eqsigma}) and (\ref{eqeta}) into
Eq.~(\ref{Lagrsigmaeta}) reproduces the original four-fermion
interactions.

The expectation values of the auxiliary fields define the chiral and
excitonic order parameters,
\begin{align}
\sigma_c
\equiv
\langle\sigma\rangle
&=
-\frac{g_c}{N}
\sum_{j=1}^{N}
\left\langle
\bar{\psi}_j\psi_j
\right\rangle ,
\label{sigmacdef}\\
\eta_c
\equiv
\langle\eta\rangle
&=
-\frac{g_e}{N}
\sum_{j=1}^{N}
\left\langle
\bar{\psi}_j i\gamma_5\psi_j
\right\rangle .
\label{etacdef}
\end{align}
The two condensates enter the quasiparticle spectrum in different
ways. The scalar field $\sigma_c$ renormalizes the explicit mass,
$m\rightarrow m+\sigma_c$, whereas $\eta_c$ produces an off-diagonal
interband gap. For spatially homogeneous condensates, the resulting
quasiparticle energy is
\begin{equation}
E_{\mathbf p}
=
\sqrt{
\mathbf p^2+
(m+\sigma_c)^2+
\eta_c^2
}.
\label{dispersion0}
\end{equation}
It is useful to define
\begin{equation}
\rho^2
\equiv
(m+\sigma_c)^2+\eta_c^2 ,
\label{rhodef}
\end{equation}
so that $E_{\mathbf p}=\sqrt{\mathbf p^2+\rho^2}$. Equation
(\ref{dispersion0}) explicitly shows that both condensates contribute
to the fermionic excitation gap. However, their physical origins are
 distinct: $\sigma_c$ belongs to the conventional scalar
mass channel, while $\eta_c$ describes interaction-induced
interband coherence and is the EI order parameter.

{}For constant background fields $\sigma_c$ and $\eta_c$, the
leading-order large-$N$ approximation reduces the problem to the
minimization of the thermodynamic potential. At temperature $T$ and
chemical potential $\mu$, the grand canonical thermodynamic potential
per unit area is
\begin{equation}
\Omega(\sigma_c,\eta_c;T,\mu)
=
-\frac{1}{\beta{\cal A}}
\ln Z(\sigma_c,\eta_c;T,\mu),
\label{Omega}
\end{equation}
where $\beta=1/T$ and ${\cal A}$ denotes the area of the planar
system. {}For fixed uniform auxiliary fields, the grand canonical
partition function is
\begin{equation}
Z(\sigma_c,\eta_c;T,\mu)
=
\int
\prod_{j=1}^{N}
{\cal D}\bar{\psi}_j{\cal D}\psi_j\,
\exp\left[
-S_E(\bar{\psi},\psi;\sigma_c,\eta_c)
\right],
\label{partfun}
\end{equation}
where the fermionic fields satisfy the antiperiodic boundary
conditions $\psi_j(\mathbf x,\tau+\beta)
= -\psi_j(\mathbf x,\tau)$.
The Euclidean action at vanishing magnetic field, but finite temperature $T=1/\beta$ and
chemical potential $\mu$, is
\begin{align}
S_E
={}&
\int_0^\beta d\tau
\int d^2x
\Bigg\{
\sum_{j=1}^{N}
\bar{\psi}_j
\left[
\gamma^0(\partial_\tau-\mu)
+i\boldsymbol{\gamma}\cdot\boldsymbol{\nabla}
+m
\right.
\nonumber \\
& \left. +\sigma_c+i\eta_c\gamma_5
\right]
\psi_j + \frac{N}{2g_c}\sigma_c^2
+\frac{N}{2g_e}\eta_c^2
\Bigg\}.
\label{action}
\end{align}
The chemical potential therefore enters as the temporal component of
a background Abelian gauge field. The fermionic part of the action is
quadratic and can be integrated out exactly. The remaining functional
of the auxiliary fields is proportional to $N$, so that the
homogeneous saddle-point approximation becomes exact in the
$N\rightarrow\infty$ limit.

At $T=\mu=0$, the unrenormalized result can be written in Euclidean
space as
\begin{align}
\frac{\Omega_0(\sigma_c,\eta_c)}{N}
={}&
\frac{\sigma_c^2}{2g_c}
+
\frac{\eta_c^2}{2g_e}
\nonumber\\
&-
2
\int^\Lambda
\frac{d^3p_E}{(2\pi)^3}
\ln
\left(
p_E^2+\rho^2
\right),
\label{Veff01}
\end{align}
where $\Lambda$ is an ultraviolet momentum cutoff and an additive
field-independent constant is understood to have been omitted.

At finite temperature and chemical potential, the temporal momentum
is replaced by the fermionic Matsubara frequencies
$ \omega_\nu=(2\nu+1)\pi T$,
$\nu\in\mathbb Z$,
and the thermodynamic potential becomes
\begin{eqnarray}
\lefteqn{
\frac{\Omega(\sigma_c,\eta_c;T,\mu)}{N}
=
\frac{\sigma_c^2}{2g_c}
+
\frac{\eta_c^2}{2g_e} }
\nonumber\\
& &-
2T
\sum_{\nu=-\infty}^{\infty}
\int^\Lambda
\frac{d^2p}{(2\pi)^2}
\ln
\left[
(\omega_\nu-i\mu)^2
+\mathbf p^2+\rho^2
\right].
\label{Veff001}
\end{eqnarray}
After performing the Matsubara sum and implementing the
renormalization prescription described in the Appendix~\ref{appA}, 
one obtains, up to a field-independent additive constant,
\begin{align}
\frac{\Omega(\sigma_c,\eta_c;T,\mu)}{N}
&=
\frac{\rho^3}{3\pi}
+
\frac{1}{2}
\left[
\frac{\sigma_c^2}{g_c}
-
\frac{(m+\sigma_c)^2}{g_\Lambda}
\right]
\nonumber\\
&+
\frac{1}{2}
\left(
\frac{1}{g_e}
-
\frac{1}{g_\Lambda}
\right)
\eta_c^2
\nonumber\\
&-
\frac{T}{\pi}
\int_0^\infty dp\,p
\Bigg[
\ln\left(
1+e^{-(E_p-\mu)/T}
\right)
\nonumber\\
&+
\ln\left(
1+e^{-(E_p+\mu)/T}
\right)
\Bigg],
\label{VeffTmu}
\end{align}
where
\begin{equation}
E_p=\sqrt{p^2+\rho^2},
\qquad
g_\Lambda=\frac{\pi}{\Lambda}.
\end{equation}
The first three terms in Eq.~(\ref{VeffTmu}) contain the vacuum
contribution and the renormalized interaction terms, whereas the last
two logarithms describe thermal fermion and antifermion excitations,
respectively.

The equilibrium condensates correspond to the global minimum of the
thermodynamic potential. They satisfy the coupled gap equations
\begin{align}
\left.
\frac{\partial\Omega}
{\partial\sigma_c}
\right|_{
\sigma_c=\bar{\sigma}_c,\,
\eta_c=\bar{\eta}_c}
&=0,
\label{gapsig}\\
\left.
\frac{\partial\Omega}
{\partial\eta_c}
\right|_{
\sigma_c=\bar{\sigma}_c,\,
\eta_c=\bar{\eta}_c}
&=0.
\label{gapeta}
\end{align}
The solutions $\bar{\sigma}_c$ and $\bar{\eta}_c$ must be obtained
self-consistently. When more than one stationary solution exists, the
corresponding values of the thermodynamic potential must be compared
in order to identify the global minimum. This comparison is essential
for distinguishing continuous transitions from first-order
transitions between competing phases.

The zero-magnetic field chiral and EI phase structures of this model
were studied in detail in Ref.~\cite{lopes}. The homogeneous
large-$N$ treatment employed here describes the formation and
thermodynamics of the fermionic excitation gap. Collective phase
fluctuations of a complex excitonic order parameter, including
possible Berezinskii--Kosterlitz--Thouless physics in strictly
two-dimensional realizations, are not included in the present
approximation. In the next section, we extend the framework above to
a constant magnetic field perpendicular to the plane.

\section{Extended Gross--Neveu model in a perpendicular magnetic field}
\label{GN_model_2+1_d_mag_field}

We now extend the zero-field formulation developed in
Sec.~\ref{GN_model_2+1_d} to a constant and homogeneous magnetic field
perpendicular to the plane. The magnetic field is treated as a fixed
external background and is coupled to the fermions through minimal
coupling. Throughout this section, we consider only the orbital effect
of the perpendicular field. Possible Zeeman splitting of the fermion
species is not included.

The Hubbard--Stratonovich Lagrangian in the presence of the external
field is obtained from Eq.~(\ref{Lagrsigmaeta}) by replacing
$\partial_\mu$ with the covariant derivative
\begin{equation}
D_\mu=\partial_\mu+i e A_\mu,
\label{covariantderivative}
\end{equation}
so that
\begin{align}
{\cal L}_B
={}&
\sum_{j=1}^{N}
\bar\psi_j
\left(
 i\gamma^\mu D_\mu
 -m-\sigma-i\eta\gamma_5
\right)
\psi_j
\nonumber\\
&-
\frac{N}{2g_c}\sigma^2
-
\frac{N}{2g_e}\eta^2 .
\label{LagrsigmaetaB}
\end{align}
Here, $e$ is the fundamental electric charge. We
choose the Landau gauge such that the external eletromagnetic gauge
field is given by
\begin{equation}
A_\mu=(0,0,Bx),
\qquad
F_{12}=B,
\label{Landaugauge}
\end{equation}
where the coordinates are ordered as $(t,x,y)$. Physical quantities
will depend only on
\begin{equation}
b\equiv |eB|,
\label{bdefinition}
\end{equation}
and are therefore independent of the particular gauge choice. 

As discussed in Appendix~\ref{appA}, it is convenient to perform the
shift
\begin{equation}
\Sigma_c\equiv m+\sigma_c .
\label{sigmashift}
\end{equation}
From this point onward, and in order to maintain the notation used in
the numerical analysis, we relabel the shifted field as
$\Sigma_c\to\sigma_c$. Thus, in the remainder of the paper,
$\sigma_c$ denotes the full scalar contribution to the fermionic gap,
including the explicit semiconductor mass. With this convention, the
field-dependent quasiparticle gap is
\begin{equation}
\rho^2\equiv\sigma_c^2+\eta_c^2,
\label{rhodefB}
\end{equation}
and the Landau-level energies are
\begin{equation}
E_n=
\sqrt{
\rho^2+2nb
},
\qquad
n=0,1,2,\ldots .
\label{Landauenergy}
\end{equation}
The excitonic condensate therefore contributes to every Landau level
through the same invariant gap combination $\rho$, while the magnetic
field controls both the spacing and the degeneracy of the levels.

{}For the four-component reducible Dirac representation employed here,
the lowest Landau level is nondegenerate with respect to the level
index, whereas all levels with $n\geq1$ have twice that degeneracy. We
accordingly define
\begin{equation}
\alpha_n=2-\delta_{n0}.
\label{alphandef}
\end{equation}
The continuum momentum contribution appearing at $B=0$ is then
replaced according to
\begin{equation}
2\int\frac{d^2p}{(2\pi)^2}\,
F(E_{\mathbf p})
\quad\longrightarrow\quad
\frac{b}{2\pi}
\sum_{n=0}^{\infty}
\alpha_n F(E_n),
\label{LandauReplacement}
\end{equation}
where $b/(2\pi)$ is the Landau degeneracy per unit area. Together with
the fermionic Matsubara sum, Eq.~(\ref{LandauReplacement}) incorporates
the complete orbital effect of the external field at leading order in
the large-$N$ expansion~\cite{Klimenko1,Klimenko2,Klimenko3,Gusynin1,Gusynin2,Kneur:2013cva}.

The ultraviolet divergence of the fermion determinant is the same as
at $B=0$. Consequently, no new magnetic field-dependent counterterms
are required: the zero-field renormalization conditions derived in
Appendix~\ref{appA} also renormalize the theory at finite $B$. The
finite magnetic contribution can be expressed compactly in terms of
the Hurwitz zeta function. One then obtains for the magnetic field
dependent thermodynamic potential the result
\begin{align}
\frac{\Omega_B}{N}
&=
\frac{\sigma_c^2}{2g_c^R}
-
{\cal H}_c\sigma_c
+
\frac{\eta_c^2}{2g_e^R}
\nonumber\\
&+
\frac{b\rho}{2\pi}
-
\frac{(2b)^{3/2}}{2\pi}
\zeta\!\left(
-\frac{1}{2},
\frac{\rho^2}{2b}
\right)
\nonumber\\
&-
\frac{bT}{2\pi}
\sum_{n=0}^{\infty}
\alpha_n
\Bigg[
\ln\!\left(
1+e^{-(E_n-\mu)/T}
\right)
\nonumber\\
&
+
\ln\!\left(
1+e^{-(E_n+\mu)/T}
\right)
\Bigg],
\label{VeffHTmu}
\end{align}
where
\begin{equation}
{\cal H}_c
\equiv
\frac{M}{g_c^R}
+
\frac{M|M|}{\pi}.
\label{Hcdefinition}
\end{equation}
The parameter $M$ is defined as the zero-temperature, zero-density,
and zero-field stationary value of the shifted scalar field on the
branch with $\eta_c=0$. The mass-renormalization condition derived in
Appendix~\ref{appA} gives equivalently
\begin{equation}
{\cal H}_c=\frac{m}{g_c},
\label{Hcmassrelation}
\end{equation}
which makes explicit that ${\cal H}_c$ acts as the external source
associated with the bare semiconductor gap.

The second line of Eq.~(\ref{VeffHTmu}) is the renormalized vacuum
contribution in the magnetic background. The two logarithms in the
last two lines describe thermally populated particle and antiparticle
Landau levels, respectively. In the limit $b\to0$, the standard
large-argument expansion of the Hurwitz zeta function, together with
the continuum limit of the Landau sum, reproduces the zero-field
thermodynamic potential. In particular, at $T=\mu=0$,
\begin{equation}
\lim_{b\to0}
\frac{\Omega_B}{N}
=
\frac{\sigma_c^2}{2g_c^R}
-
{\cal H}_c\sigma_c
+
\frac{\eta_c^2}{2g_e^R}
+
\frac{\rho^3}{3\pi},
\label{veffT0}
\end{equation}
in agreement with the shifted and renormalized form of the result
derived in Sec.~\ref{GN_model_2+1_d} and Appendix~\ref{appA}.

The equilibrium condensates are determined by minimizing
Eq.~(\ref{VeffHTmu}) with respect to both order parameters. Using
\begin{equation}
\frac{\partial}{\partial a}
\zeta(s,a)
=
-s\,\zeta(s+1,a),
\label{zetaderivative}
\end{equation}
the scalar gap equation becomes
\begin{align}
0
={}&
\frac{\sigma_c}{g_c^R}
-{\cal H}_c
+
\frac{b}{2\pi}
\frac{\sigma_c}{\rho}
-
\frac{\sqrt{b}}{\sqrt{2}\,\pi}
\sigma_c
\zeta\!\left(
\frac{1}{2},
\frac{\rho^2}{2b}
\right)
\nonumber\\
&+
\frac{b}{2\pi}
\sum_{n=0}^{\infty}
\alpha_n
\left[
 f(E_n-\mu)
+
 f(E_n+\mu)
\right]
\frac{\sigma_c}{E_n},
\label{gap_equations_with_B_sigma}
\end{align}
whereas the excitonic gap equation is
\begin{align}
0
={}&
\frac{\eta_c}{g_e^R}
+
\frac{b}{2\pi}
\frac{\eta_c}{\rho}
-
\frac{\sqrt{b}}{\sqrt{2}\,\pi}
\eta_c
\zeta\!\left(
\frac{1}{2},
\frac{\rho^2}{2b}
\right)
\nonumber\\
&+
\frac{b}{2\pi}
\sum_{n=0}^{\infty}
\alpha_n
\left[
 f(E_n-\mu)
+
 f(E_n+\mu)
\right]
\frac{\eta_c}{E_n}.
\label{gap_equations_with_B_eta}
\end{align}
Here,
\begin{equation}
f(x)=\frac{1}{e^{x/T}+1}
\label{FermiFunction}
\end{equation}
is the Fermi--Dirac distribution. Equations
(\ref{gap_equations_with_B_sigma}) and
(\ref{gap_equations_with_B_eta}) must be solved simultaneously, and
all stationary solutions must be compared through
Eq.~(\ref{VeffHTmu}) in order to identify the global thermodynamic
minimum. The solution $\eta_c=0$ is always a stationary branch of the
excitonic equation. A nonzero EI solution exists when the expression
multiplying $\eta_c$ in Eq.~(\ref{gap_equations_with_B_eta}) admits a
zero and the corresponding stationary point is thermodynamically
favored.

The magnetic field affects the condensates through two related
mechanisms. It discretizes the quasiparticle spectrum and introduces
the degeneracy factor $b/(2\pi)$, thereby enhancing the spectral weight
of the low-lying Landau levels. At the same time, it changes the
thermal occupation of these levels. At sufficiently low temperature
and density, the first effect generally favors interaction-induced gap
formation and underlies magnetic catalysis. At finite $T$ or $\mu$,
however, Landau-level occupation can produce nonmonotonic and reentrant
behavior. The actual response of the EI order parameter, therefore,
follows from the coupled minimization problem rather than from the
magnetic dependence of the single-particle spectrum alone.

\subsection{Zero-temperature limit at finite chemical potential}

The $T=0$ limit is particularly useful for understanding the
Landau-level structure of the phase diagram. For $\mu\geq0$, the
thermal terms obey
\begin{align}
\lim_{T\to0}
T\ln\!\left(
1+e^{-(E_n-\mu)/T}
\right)
&=
(\mu-E_n)\,
\theta(\mu-E_n),
\label{T0limitparticle}\\
\lim_{T\to0}
T\ln\!\left(
1+e^{-(E_n+\mu)/T}
\right)
&=0.
\label{T0limitantiparticle}
\end{align}
The thermodynamic potential consequently reduces to
\begin{align}
\frac{\Omega_B(
\sigma_c,\eta_c;T=0,\mu,b)}{N}
&=
\frac{\sigma_c^2}{2g_c^R}
-
{\cal H}_c\sigma_c
+
\frac{\eta_c^2}{2g_e^R}
\nonumber\\
&+
\frac{b\rho}{2\pi}
-
\frac{(2b)^{3/2}}{2\pi}
\zeta\!\left(
-\frac{1}{2},
\frac{\rho^2}{2b}
\right)
\nonumber\\
&-
\frac{b}{2\pi}
\sum_{n=0}^{\infty}
\alpha_n
(\mu-E_n)
\theta(\mu-E_n).
\label{VeffHT0muTheta}
\end{align}
Only levels satisfying $E_n\leq\mu$ contribute to the density-dependent
part. When $\mu>\rho$, the highest occupied Landau level is therefore
\begin{equation}
n_{\rm max}
=
{\rm Int}\left\lfloor
\frac{\mu^2-\rho^2}{2b}
\right\rfloor,
\qquad
\mu>\rho,
\label{nmax}
\end{equation}
where ${\rm Int}\lfloor x\rfloor$ denotes the greatest integer not exceeding
$x$. {}For $\mu\leq\rho$, no positive-energy Landau level is occupied and
the density-dependent sum vanishes. Hence, Eq.~(\ref{VeffHT0muTheta}) can be written as
\begin{align}
\frac{\Omega_B(
\sigma_c,\eta_c;T=0,\mu,b)}{N}
&=
\frac{\sigma_c^2}{2g_c^R}
-
{\cal H}_c\sigma_c
+
\frac{\eta_c^2}{2g_e^R}
\nonumber\\
&+
\frac{b\rho}{2\pi}
-
\frac{(2b)^{3/2}}{2\pi}
\zeta\!\left(
-\frac{1}{2},
\frac{\rho^2}{2b}
\right)
\nonumber\\
&-
\frac{b}{2\pi}
\sum_{n=0}^{n_{\rm max}}
\alpha_n
(\mu-E_n),
\qquad
\mu>\rho,
\label{VeffHT0mu}
\end{align}
with the last line omitted when $\mu\leq\rho$. The corresponding
zero-temperature gap equations follow directly from
Eqs.~(\ref{gap_equations_with_B_sigma}) and
(\ref{gap_equations_with_B_eta}) through the replacements
\begin{equation}
f(E_n-\mu)\to\theta(\mu-E_n),
\qquad
f(E_n+\mu)\to0.
\label{T0FermiReplacement}
\end{equation}
The discontinuous change of $n_{\rm max}$ as $\mu$ or $B$ is varied is
the origin of the Landau-level oscillations and of the multiple
competing stationary branches that may appear in the finite-density
phase structure~\cite{Ramos:2013aia,Kneur:2013cva}.

\subsection{Hall conductivity}

The Landau-level filling pattern also has a direct transport
signature. Within the clean, dissipationless approximation used here,
the Hall conductivity is related to the equilibrium fermion number
density by~\cite{Ramos:2013aia,kittel}
\begin{equation}
\Sigma_{xy}
=
\operatorname{sgn}(eB)
\frac{e^2}{b}\,n,
\label{hallcond}
\end{equation}
where the total number density is obtained from the thermodynamic
potential evaluated at its global minimum,
\begin{equation}
n
=
-
\left.
\frac{\partial \Omega}
{\partial\mu}
\right|_{
\sigma_c=\bar\sigma_c,\,
\eta_c=\bar\eta_c}.
\label{numberdensitydefinition}
\end{equation}
Because the equilibrium fields satisfy the gap equations, their
implicit $\mu$ dependence does not contribute to the first derivative
in Eq.~(\ref{numberdensitydefinition}). Differentiating
Eq.~(\ref{VeffHTmu}) gives
\begin{equation}
n
=
\frac{Nb}{2\pi}
\sum_{n=0}^{\infty}
\alpha_n
\left[
 f(\bar E_n-\mu)
-
 f(\bar E_n+\mu)
\right],
\label{numberdensity}
\end{equation}
where
\begin{equation}
\bar E_n
=
\sqrt{
2nb+\bar\sigma_c^{\,2}+\bar\eta_c^{\,2}
}
\label{equilibriumLandauenergy}
\end{equation}
is evaluated at the global minimum. It is convenient to define the
normalized Hall conductivity
\begin{equation}
\bar\Sigma_{xy}
\equiv
\frac{2\pi}{Ne^2}\,
\sigma_{xy}.
\label{normalizedHall}
\end{equation}
At finite temperature,
\begin{equation}
\bar\Sigma_{xy}
=
\operatorname{sgn}(eB)
\sum_{n=0}^{\infty}
\alpha_n
\left[
 f(\bar E_n-\mu)
-
 f(\bar E_n+\mu)
\right].
\label{HallFiniteT}
\end{equation}
{}For $T=0$, $\mu\geq0$, and $eB>0$, this expression reduces to
\begin{equation}
\bar\Sigma_{xy}
=
\left[
1+2
{\rm Int}\left\lfloor
\frac{
\mu^2-\bar\sigma_c^{\,2}-\bar\eta_c^{\,2}
}{2b}
\right\rfloor
\right]
\theta\!\left(
\mu-
\sqrt{
\bar\sigma_c^{\,2}+\bar\eta_c^{\,2}
}
\right).
\label{HallT0}
\end{equation}
Equation~(\ref{HallT0}) displays the usual plateau structure associated
with successive Landau-level filling. In the present model, however,
the plateau thresholds are not fixed solely by the externally applied
field: they also depend self-consistently on the chiral and excitonic
condensates. A continuous or discontinuous change of the EI gap can
therefore shift the onset of a finite Hall response or generate an
additional jump in $\Sigma_{xy}$. This correlation between the Hall
conductivity and the appearance of the EI condensate will be
examined numerically next, in Sec.~\ref{numerical_results}.

\section{Numerical results}
\label{numerical_results}

We now present the numerical solutions of the coupled gap equations,
Eqs.~(\ref{gap_equations_with_B_sigma}) and
(\ref{gap_equations_with_B_eta}), and analyze the resulting excitonic
and scalar condensates, phase boundaries, and Hall response. At each
point in the parameter space, all stationary solutions were compared using
the thermodynamic potential in Eq.~(\ref{VeffHTmu}), so that the curves
shown below correspond to the global thermodynamic minimum.

{}Following Ref.~\cite{lopes}, we choose the renormalized couplings in a
region that supports an EI ground state in the semiconductor regime,
namely, we use the following representative values in our numerical examples\footnote{As explained in Ref.~\cite{lopes}, we require the hierarchy $g_e > g_\Lambda > g_c$, regardless of specific numerical values. Different parameter choices merely renormalize the scale of the critical parameters. Therefore, the qualitative thermodynamic behavior of the system remains unchanged under the choice of parameters made here.},
\begin{equation}
 g_e^R=-11\,\sigma_0^{-1},
 \qquad
 g_c^R=9\,\sigma_0^{-1}.
 \label{numericalcouplings}
\end{equation}
Unless otherwise stated, we set $M=0.24\sigma_0$. Here, $\sigma_0$ is
the value of the scalar condensate in the EI phase and is used to
normalize all dimensionful quantities. For the positive magnetic fields
considered in the {}Figures, the variable $b=|eB|$ introduced in
Eq.~(\ref{bdefinition}) reduces to $b=eB$.

An important property of the coupled gap equations is that the scalar
condensate is fixed throughout the EI phase. As shown in
Appendix~\ref{AppB}, whenever $\bar\eta_c\neq0$ one obtains
\begin{equation}
 \bar\sigma_c=\sigma_0
 =
 \frac{
 g_e^R M\left(g_c^R|M|+\pi\right)
 }{
 \pi\left(g_e^R-g_c^R\right)
 }.
 \label{sigmacEIconstant}
\end{equation}
This relation remains valid at finite $T$, $\mu$, and $B$. Consequently,
the magnetic field dependence of the total quasiparticle gap inside the
EI phase is carried by $\bar\eta_c$, while $\bar\sigma_c$ changes only
after the EI condensate has vanished.

\subsection{Thermal and magnetic field dependence at $\mu=0$}

\begin{figure}[!htb]
 \centering
 \includegraphics[width=0.91\linewidth]{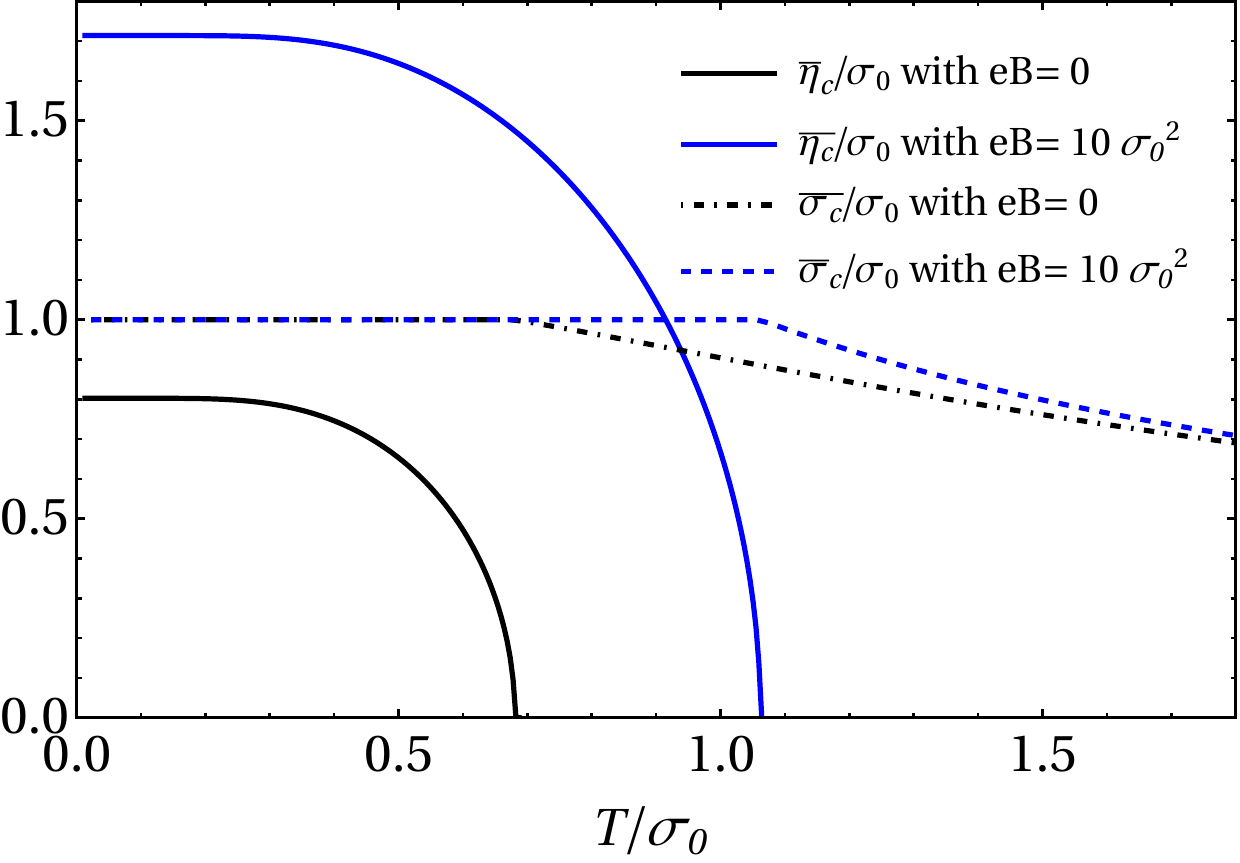}
 \caption{Normalized EI condensate (solid curves) and
 scalar condensate (dashed curves) as functions of temperature for
 $eB=0$ and $eB=10\sigma_0^2$, with $M=0.24\sigma_0$. The magnetic
 field enhances the EI gap and raises its continuous transition
 temperature. The scalar condensate remains equal to $\sigma_0$ while
 $\bar\eta_c\neq0$ and becomes temperature dependent only in the
 non-EI phase.}
 \label{fig1}
\end{figure}

{}Figure~\ref{fig1} shows the temperature dependence of the two
condensates for $M=0.24\sigma_0$. At $B=0$ and $T=0$, the equilibrium
solution has $\bar\eta_c\simeq0.8\sigma_0$ and
$\bar\sigma_c=\sigma_0$. The finite value of $\bar\sigma_c$ should not
be interpreted as an independent spontaneous transition in the present
semiconductor regime, because the explicit mass parameter already
breaks the corresponding chiral symmetry. The genuine EI order is
identified by $\bar\eta_c\neq0$.
As the temperature increases, the EI condensate decreases continuously
and vanishes at
\begin{equation}
 T_c(B=0)\simeq0.67\sigma_0.
\end{equation}
The transition is continuous, and, hence, second order within the
large-$N$ mean-field approximation. Equation~(\ref{sigmacEIconstant})
explains why the scalar condensate remains exactly constant up to this
transition. Above $T_c$, where $\bar\eta_c=0$, the scalar condensate
becomes temperature dependent and decreases smoothly because the
explicit mass prevents a true restoration of the corresponding
symmetry. The zero-field behavior agrees with Ref.~\cite{lopes}.

{}For $eB=10\sigma_0^2$, both the zero-temperature EI gap and its thermal
stability are enhanced. In this case,
$\bar\eta_c(T=0)\simeq1.72\sigma_0$, and the EI condensate survives up
to
\begin{equation}
 T_c(eB=10\sigma_0^2)\simeq1.05\sigma_0.
\end{equation}
The increase of both $\bar\eta_c(0)$ and $T_c$ demonstrates magnetic
catalysis in the excitonic channel. By contrast, the scalar condensate
is pinned to $\sigma_0$ while the EI solution is present. Magnetic
catalysis of the scalar channel becomes visible only on the
$\bar\eta_c=0$ branch.

\begin{figure}[!htb]
 \centering
 \includegraphics[width=0.91\linewidth]{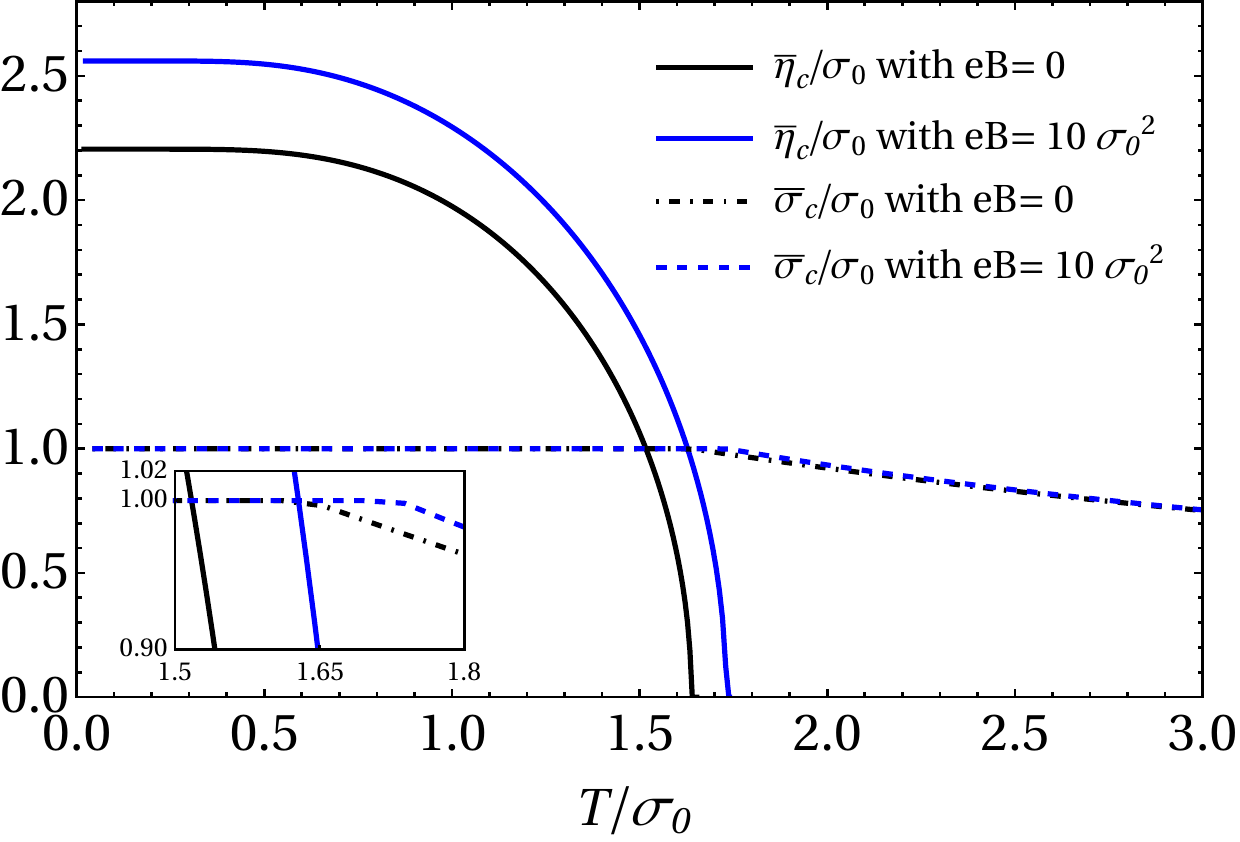}
 \caption{Normalized EI condensate (solid curves) and
 scalar condensate (dashed curves) as functions of temperature for
 $M=0.15\sigma_0$, at $eB=0$ and $eB=10\sigma_0^2$. The smaller bare
 gap favors the EI phase, while the magnetic field further increases
 the condensate and its critical temperature. The inset enlarges the
 behavior of the scalar condensate near the EI transition.}
 \label{fig2}
\end{figure}

The same qualitative behavior is obtained for the smaller explicit gap
parameter $M=0.15\sigma_0$, as shown in {}Fig.~\ref{fig2}. Reducing $M$
weakens the bare semiconductor gap relative to the attractive
excitonic channel, thereby enlarging the EI region already at $B=0$.
The perpendicular field further increases the EI condensate and its
critical temperature. For $eB=10\sigma_0^2$, the continuous EI
transition occurs near $T_c\simeq1.73\sigma_0$. The inset makes clear
that the scalar condensate remains pinned until the EI order disappears
and then crosses smoothly to the non-EI branch.

\begin{figure}[!htb]
 \centering
 \includegraphics[width=0.91\linewidth]{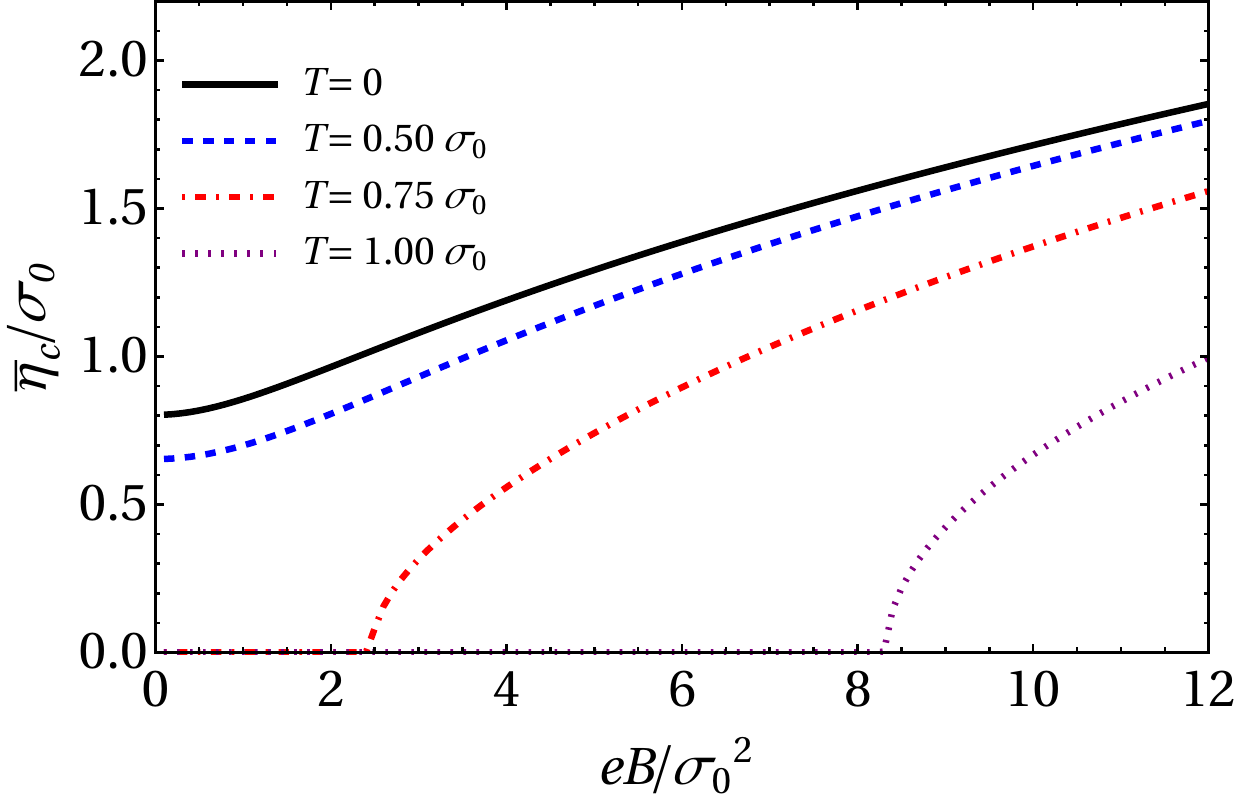}
 \caption{Normalized EI condensate as a function of the
 perpendicular magnetic field for several temperatures, with
 $M=0.24\sigma_0$. The monotonic enhancement of $\bar\eta_c$ is the
 excitonic analogue of magnetic catalysis. At temperatures above the
 zero-field critical temperature, a sufficiently strong field induces
 the EI phase.}
 \label{fig3}
\end{figure}

The magnetic field dependence of the EI condensate is displayed more
directly in {}Fig.~\ref{fig3}. For every temperature shown,
$\bar\eta_c$ is a nondecreasing function of $eB$. At $T=0$ and
$T=0.50\sigma_0$, the system is already in the EI phase at zero field,
and the applied field continuously strengthens the condensate. At
$T=0.75\sigma_0$ and $T=1.00\sigma_0$, by contrast, the zero-field
state lies above its EI transition temperature. A sufficiently strong
field then induces the condensate, with approximate onset fields
$eB\simeq2\sigma_0^2$ and $eB\simeq8\sigma_0^2$, respectively. Thus,
the field not only enhances an existing EI state but may also drive a
normal to EI transition at fixed temperature.

\begin{figure}[!htb]
 \centering
 \includegraphics[width=0.91\linewidth]{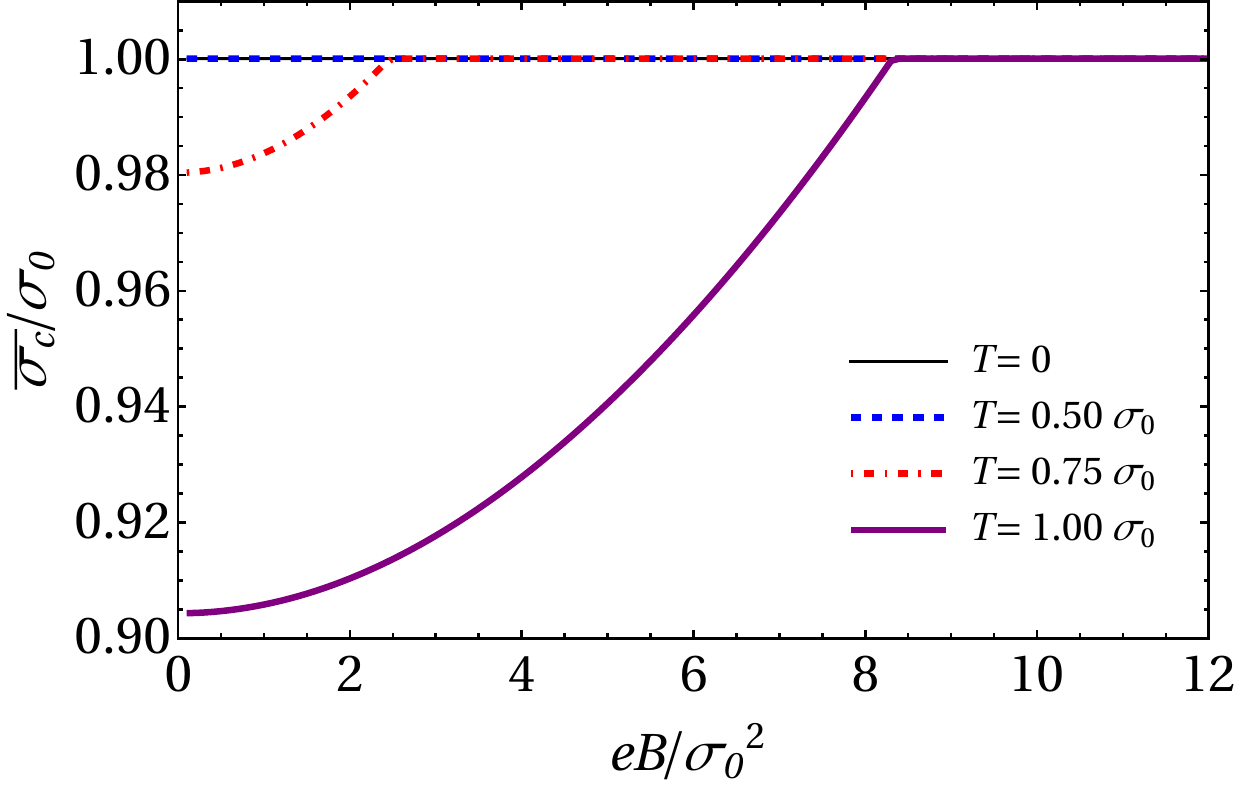}
 \caption{Normalized scalar condensate as a function of
 the perpendicular magnetic field for several temperatures. In the EI
 phase, $\bar\sigma_c=\sigma_0$ and is independent of $B$. Above the
 zero-field EI transition, the scalar condensate initially exhibits
 magnetic catalysis; it becomes pinned to $\sigma_0$ when the
 field-induced EI condensate appears.}
 \label{fig4}
\end{figure}
{}Figure~\ref{fig4} shows the corresponding scalar condensate. Whenever
the system is in the EI phase, Eq.~(\ref{sigmacEIconstant}) enforces
$\bar\sigma_c/\sigma_0=1$, independently of $T$ and $B$. This explains
the coincident $T=0$ and $T=0.50\sigma_0$ curves. For temperatures
above the zero-field EI transition, the system initially lies on the
$\bar\eta_c=0$ branch, where $\bar\sigma_c$ grows with $B$ through the
usual scalar magnetic-catalysis mechanism. Once the field-induced EI
transition is reached, $\bar\sigma_c$ attains $\sigma_0$ and becomes
constant. The behavior in {}Figs.~\ref{fig3} and \ref{fig4} therefore
exhibits a transfer between the two competing channels: the magnetic
field first enhances the scalar gap on the non-EI branch and, after the
EI state becomes thermodynamically favored, the further field
dependence is carried by $\bar\eta_c$.

In the remainder of this section, unless otherwise stated, we use
$M=0.24\sigma_0$.

\subsection{Density-driven transition at $T=0$}

\begin{figure}[!htb]
 \centering
 \includegraphics[width=0.91\linewidth]{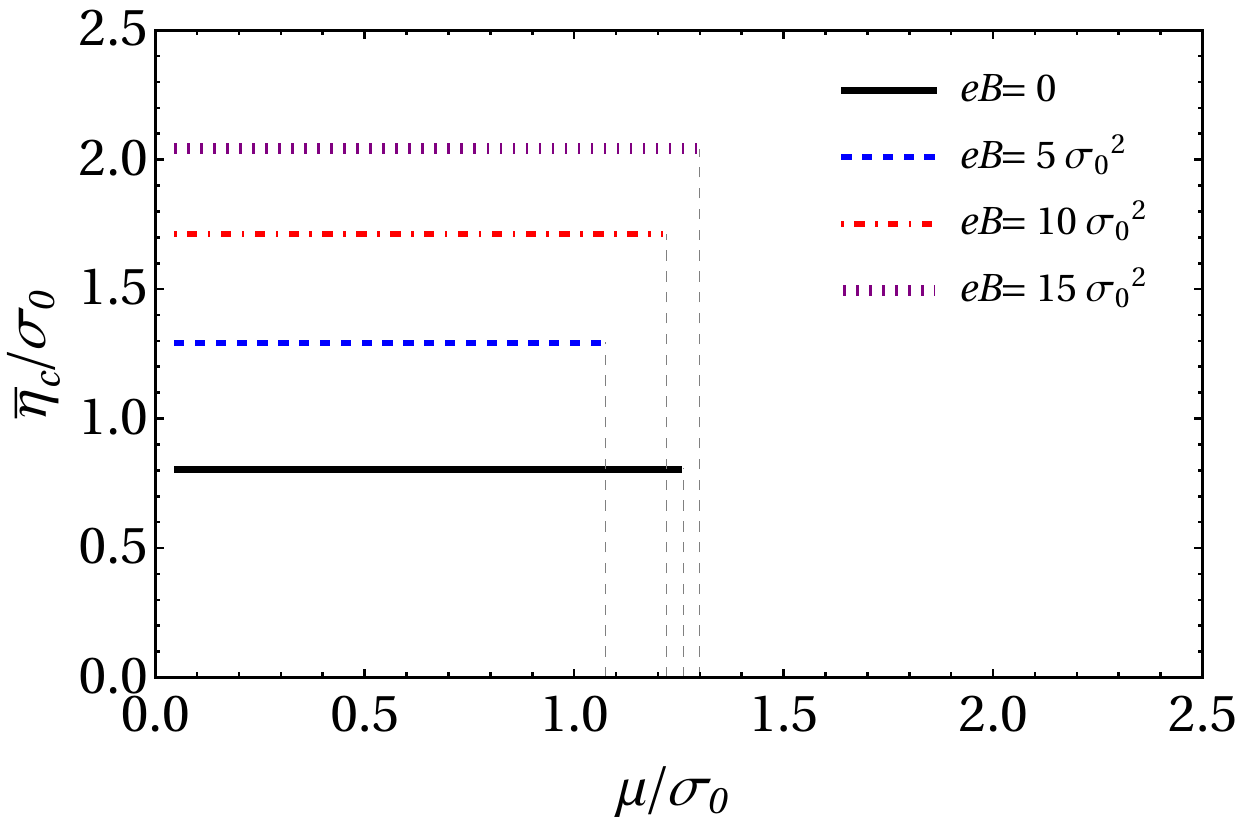}
 \caption{Normalized EI condensate as a function of
 chemical potential at $T=0$ for several magnetic fields. For every
 field shown, the EI condensate terminates through a single first-order
 transition. The critical chemical potential depends nonmonotonically
 on $B$ because of the changing Landau-level occupation.}
 \label{fig5}
\end{figure}

{}Figure~\ref{fig5} displays the EI condensate as a function of chemical
potential at zero temperature. For each magnetic field, the condensate
is nearly independent of $\mu$ throughout the EI phase and then
vanishes discontinuously at a critical value $\mu_c(B)$. The transition
is therefore first order. At zero field,
$\mu_c\simeq1.28\sigma_0$. The field dependence is nonmonotonic: for
example, the transition moves to approximately
$\mu_c\simeq1.11\sigma_0$ at $eB=5\sigma_0^2$, but shifts back toward
larger chemical potential at stronger fields.

This behavior does not contradict the monotonic magnetic enhancement
of $T_c$ found above. At finite density, the field has two competing
consequences. It enhances the interaction-driven gap through the
low-lying Landau levels, but it also changes the sequence of occupied
levels as $\mu$ is varied. The latter effect produces oscillations and
nonmonotonicity in the density-driven transition line.

\begin{figure}[!htb]
 \centering
 \includegraphics[width=0.91\linewidth]{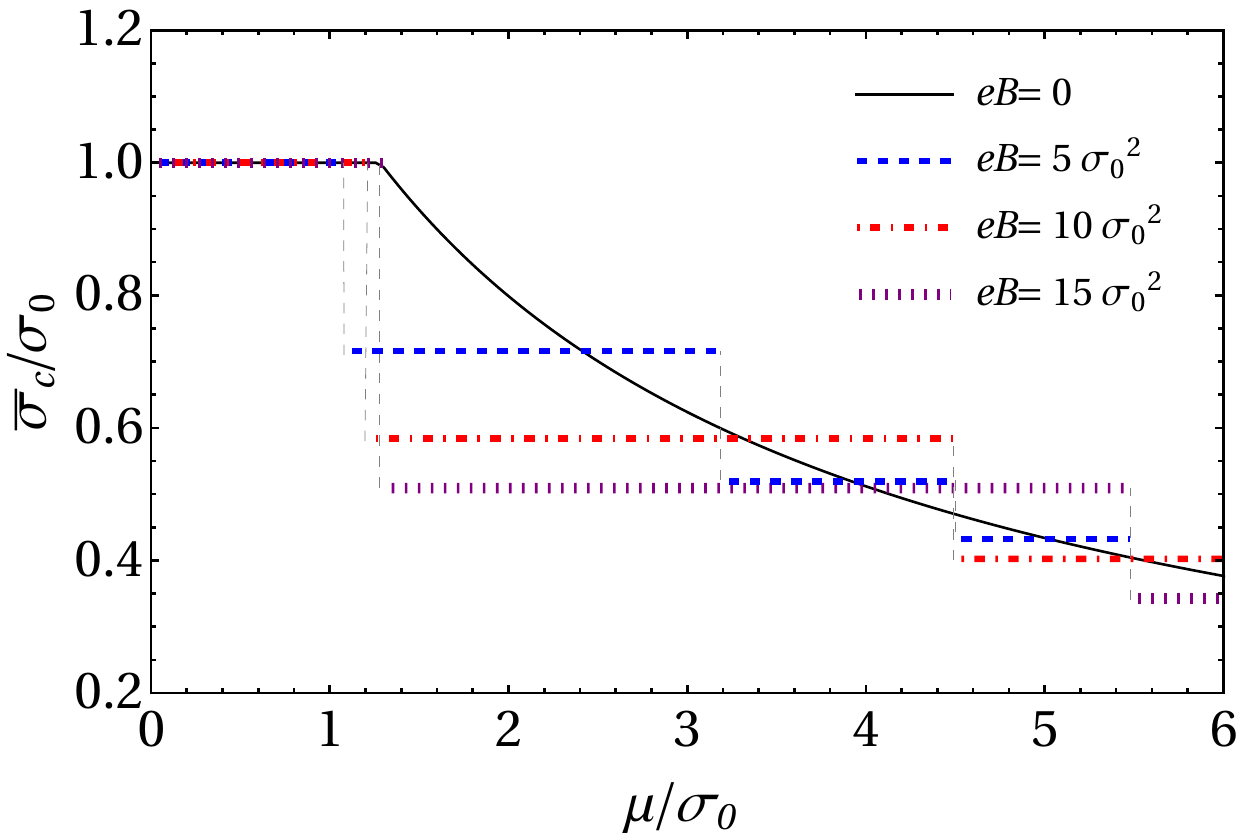}
 \caption{Normalized scalar condensate as a function of
 chemical potential at $T=0$ for several magnetic fields. The scalar
 condensate remains equal to $\sigma_0$ in the EI phase. After the EI
 transition, Landau-level occupation produces a sequence of
 discontinuous changes whose separation increases with the magnetic
 field.}
 \label{fig6}
\end{figure}

The corresponding scalar condensate is shown in {}Fig.~\ref{fig6}. It
remains fixed at $\sigma_0$ as long as $\bar\eta_c\neq0$, and changes
only after the first-order EI transition. At $B=0$, the post-transition
scalar condensate decreases smoothly with increasing $\mu$. For
$B\neq0$, it instead exhibits a sequence of discontinuous changes.
These structures arise when the global minimum readjusts as successive
Landau levels become occupied. Increasing the field increases the
Landau-level spacing, so that fewer discontinuities occur over a fixed
interval of chemical potential and the intervals between them become
wider. Similar Landau-level-induced structures have been found for the
scalar transition of the $(2+1)$-dimensional GN model in
Refs.~\cite{Ramos:2013aia,Kneur:2006ht}.

\subsection{Magnetic field dependence of the EI phase boundaries}

\begin{figure}[!htb]
 \centering
 \includegraphics[width=0.91\linewidth]{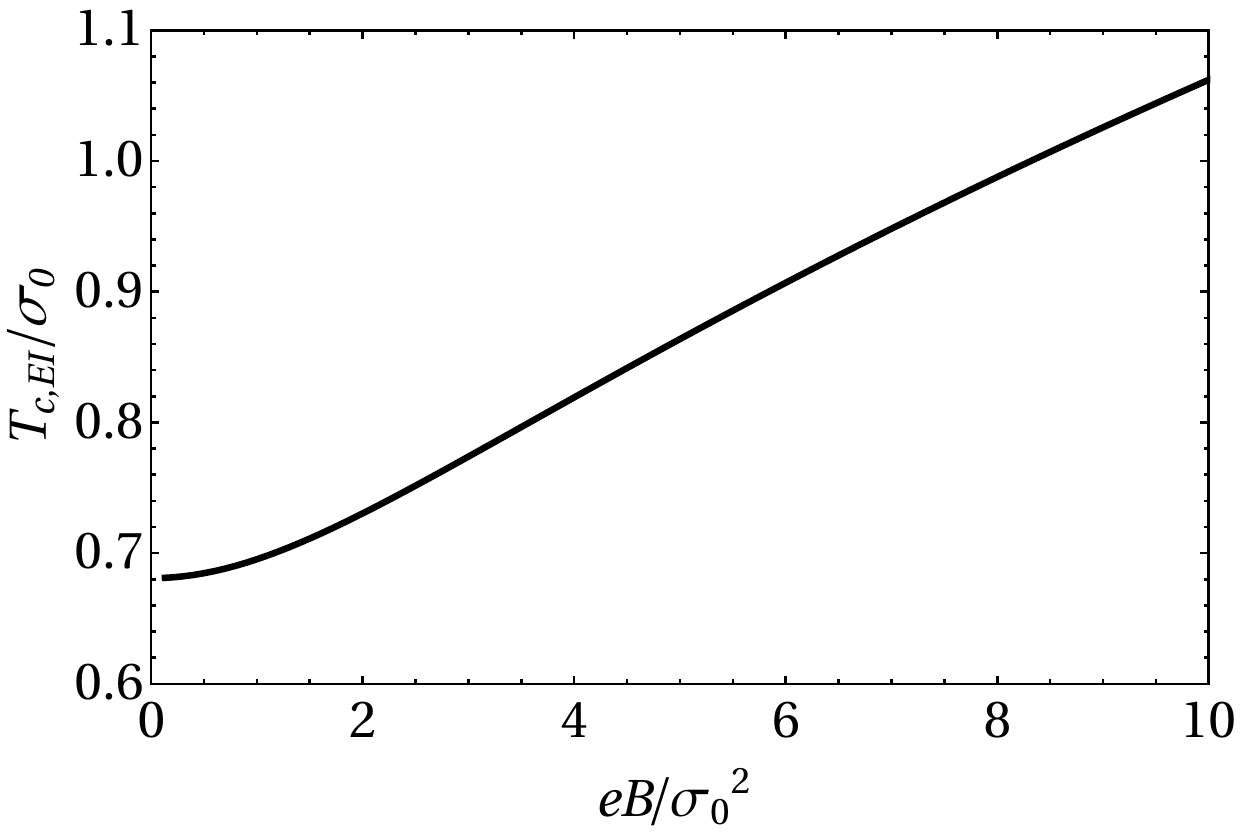}
 \caption{EI critical temperature at $\mu=0$ as a
 function of the perpendicular magnetic field. The monotonic increase
 of $T_c$ provides a direct characterization of magnetic catalysis in
 the excitonic channel.}
 \label{fig7}
\end{figure}

\begin{figure}[!htb]
 \centering
 \includegraphics[width=0.91\linewidth]{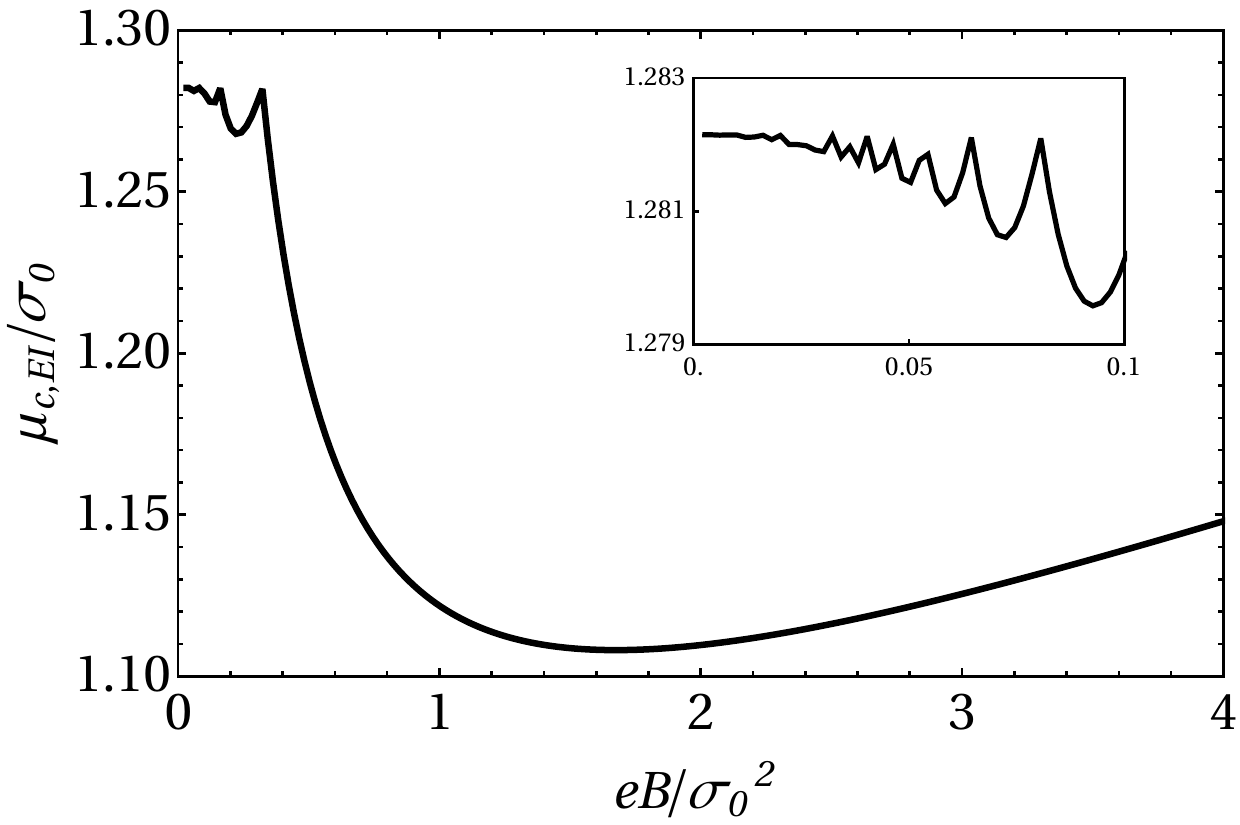}
 \caption{Critical chemical potential of the EI phase
 at $T=0$ as a function of magnetic field. The inset resolves the
 weak-field oscillations caused by changes in Landau-level occupation.
 At intermediate field, $\mu_c$ reaches a broad minimum and then rises
 on the strong-field magnetic-catalysis branch.}
 \label{fig8}
\end{figure}

The zero-density critical line provides the clearest measure of
magnetic catalysis. {}Figure~\ref{fig7} shows that $T_c$ increases
monotonically with $eB$. In the present effective theory, this behavior
follows from the enhancement of the low-energy spectral weight by
Landau quantization. In the conventional exciton language, it is also
consistent with the magnetic confinement of the relative electron-hole
motion and the associated increase of the binding energy discussed in
Refs.~\cite{Elliott,fenton,Yafet}. The two descriptions emphasize
complementary aspects of the same tendency: a perpendicular field
stabilizes the EI state against thermal fluctuations.

At $T=0$, the critical chemical potential has a more intricate field
dependence, as shown in {}Fig.~\ref{fig8}. In the weak-field region,
$eB/\sigma_0^2\lesssim0.3$, $\mu_c$ exhibits small oscillations,
shown in the inset. These are the analogue of de Haas--van Alphen-type
structures and originate from the discrete changes in the highest
occupied Landau level. Beyond this oscillatory region, $\mu_c$
decreases to a broad minimum,
$\mu_c/\sigma_0\simeq1.11$, near
$eB/\sigma_0^2\simeq1.7$, and subsequently increases. The increasing
strong-field branch reflects the eventual dominance of magnetic
catalysis over the Landau-level filling effect.

\begin{figure}[!htb]
 \centering
 \includegraphics[width=0.91\linewidth]{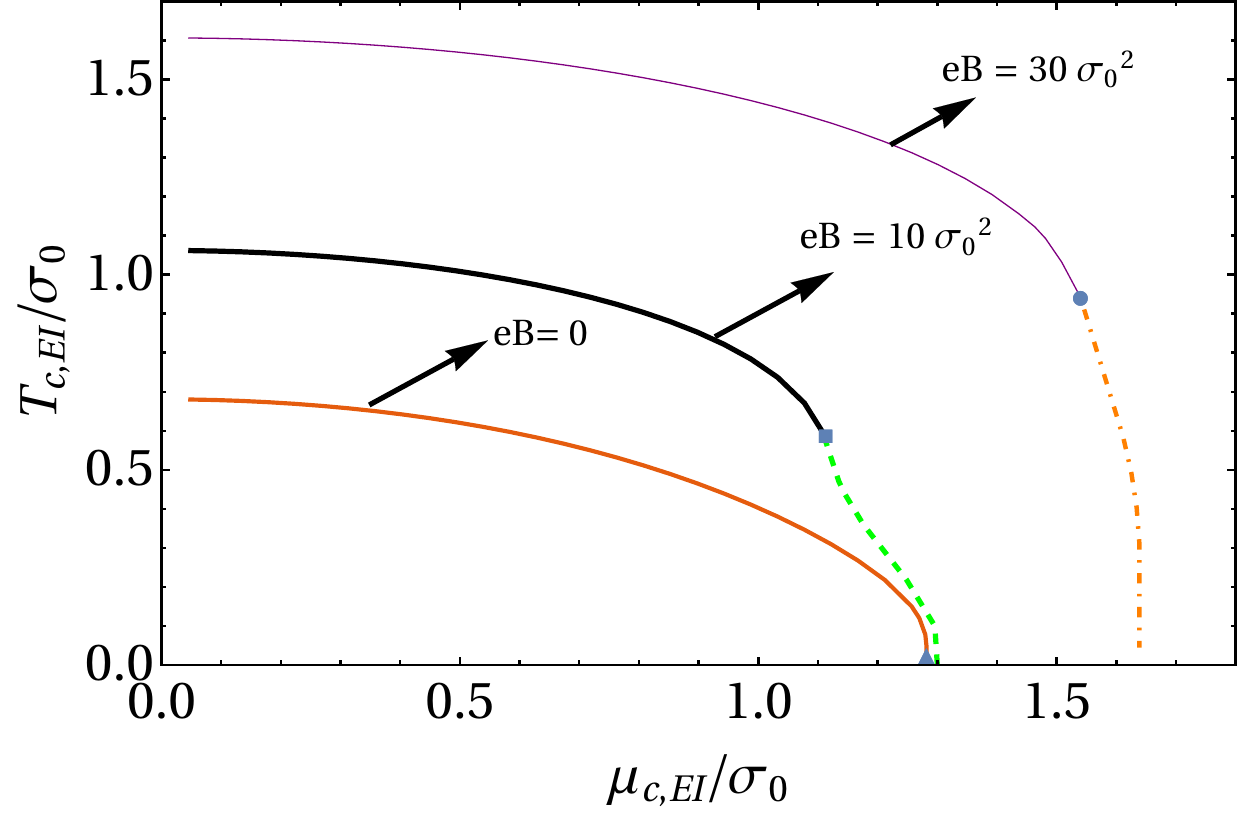}
 \caption{EI phase boundaries in the
 $T$--$\mu$ plane for $eB=0$, $10\sigma_0^2$, and
 $30\sigma_0^2$. Solid curves denote continuous transitions, while
 dashed or dash-dotted curves denote first-order transitions. The
 magnetic field expands the EI region and shifts the tricritical point
 to finite and progressively larger temperatures.}
 \label{fig9}
\end{figure}

The complete $T$--$\mu$ phase boundaries for three fixed fields are
shown in {}Fig.~\ref{fig9}. At $B=0$, the finite-temperature boundary is
continuous and terminates at the first-order zero-temperature
transition near $(\mu_c,T_c)=(1.28\sigma_0,0)$. For
$eB=10\sigma_0^2$, the EI region expands toward higher temperature and
the point at which the transition changes order moves to approximately
$(\mu_{\rm tri},T_{\rm tri}) \simeq (1.11\sigma_0,0.58\sigma_0)$.
{}For $eB=30\sigma_0^2$, it moves further to $(\mu_{\rm tri},T_{\rm tri})
 \simeq (1.54\sigma_0,0.94\sigma_0)$.

Since a continuous transition line meets a first-order line at these
points, they are tricritical points within the present mean-field phase
diagram. Thus, the magnetic field not only enlarges the EI domain but
also increases the portion of the boundary governed by a first-order
transition.

\subsection{Hall response and its correlation with EI order}

Because the opening of a spectral gap alone is not an unambiguous
signature of excitonic order, it is useful to examine a transport
observable that is sensitive to both the self-consistent gap and the
Landau-level structure. We therefore evaluate the Hall conductivity
from Eqs.~(\ref{HallFiniteT}) and (\ref{HallT0}). The discussion below
concerns the normalized quantity $\bar\Sigma_{xy}$ defined in
Eq.~(\ref{normalizedHall}).
\begin{figure}[!htb]
 \centering
 \includegraphics[width=0.91\linewidth]{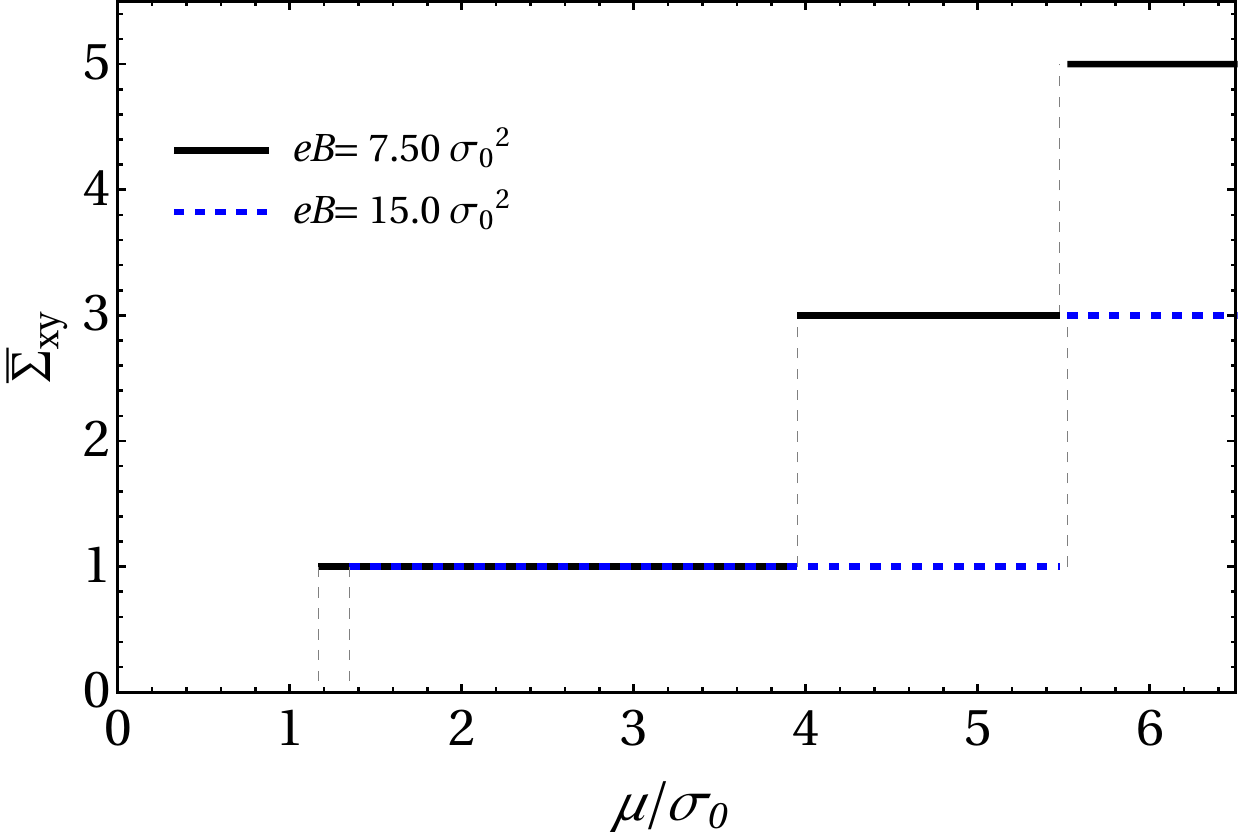}
 \caption{Normalized Hall conductivity at $T=0$ as a
 function of chemical potential for $eB=7.5\sigma_0^2$ and
 $eB=15\sigma_0^2$. Increasing the field enlarges the Landau-level
 spacing and therefore broadens the plateaus, reducing the number of
 visible steps over a fixed interval of $\mu$.}
 \label{fig10}
\end{figure}

{}Figure~\ref{fig10} shows $\bar\Sigma_{xy}$ as a function of chemical
potential at $T=0$ for two magnetic fields. The Hall response vanishes
as long as the chemical potential lies below the lowest positive-energy
Landau level,
\begin{equation}
 \mu<E_0
 =
 \sqrt{\bar\sigma_c^{\,2}+\bar\eta_c^{\,2}}.
 \label{Hallthresholdresults}
\end{equation}
Once this threshold is crossed, successive Landau levels become
occupied and the Hall conductivity develops the odd-integer sequence
implied by Eq.~(\ref{HallT0}). A stronger field increases the
Landau-level spacing. Consequently, the plateaus become wider and fewer
steps are visible over the same chemical-potential interval.

\begin{figure}[!htb]
 \centering
 \includegraphics[width=0.91\linewidth]{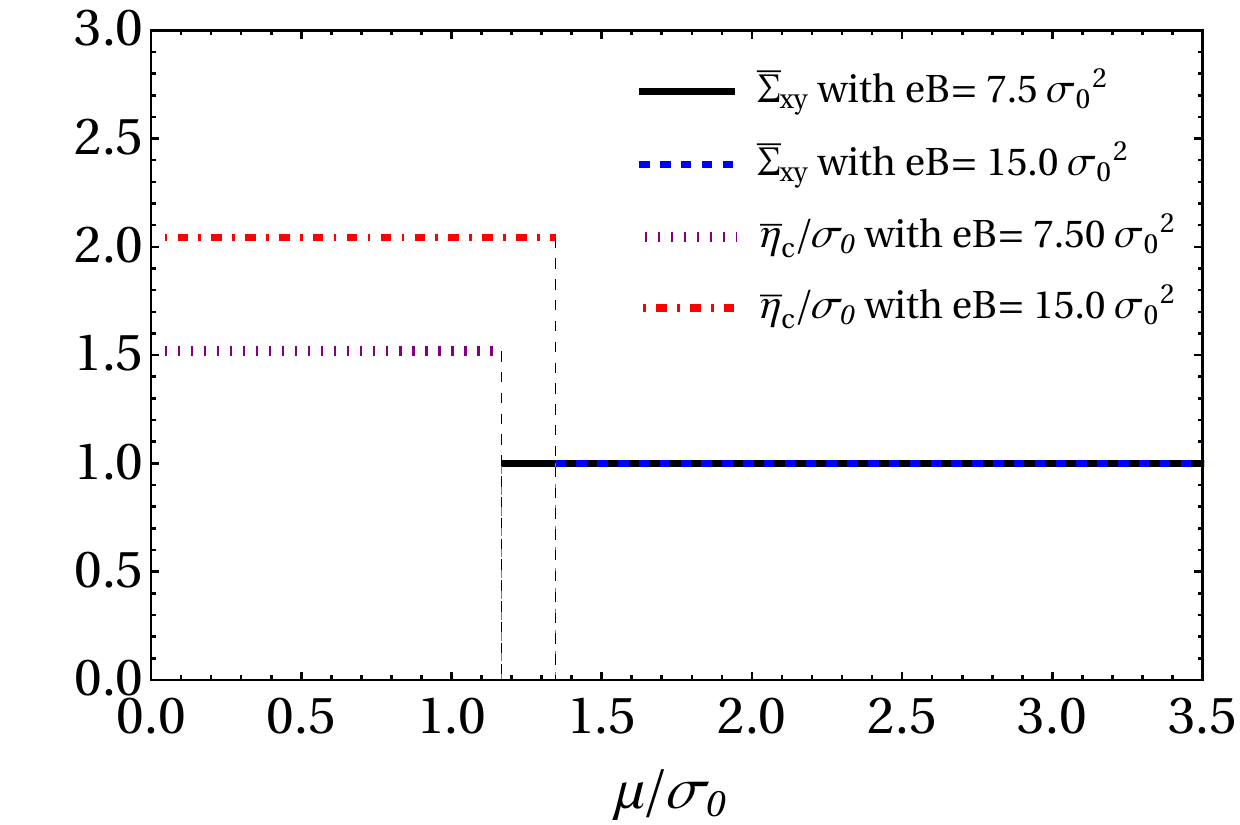}
 \caption{Normalized Hall conductivity and EI
 condensate as functions of chemical potential at $T=0$, for
 $eB=7.5\sigma_0^2$ and $eB=15\sigma_0^2$. For the parameters shown,
 the first nonzero Hall plateau appears at the same first-order
 transition at which the EI condensate vanishes.}
 \label{fig11}
\end{figure}

\begin{center}
\begin{figure}[!htb]
\subfigure[]{\includegraphics[width=7.5cm]{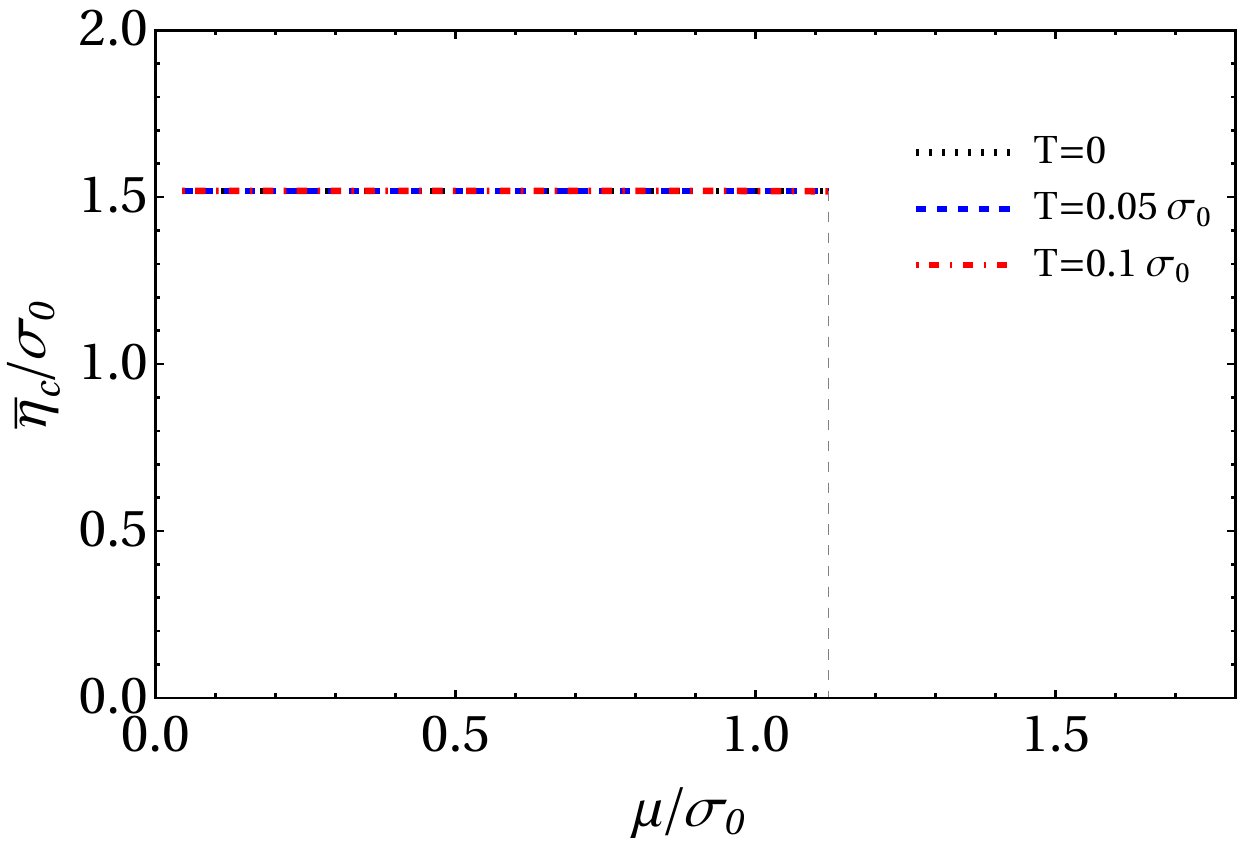}}
\subfigure[]{\includegraphics[width=7.5cm]{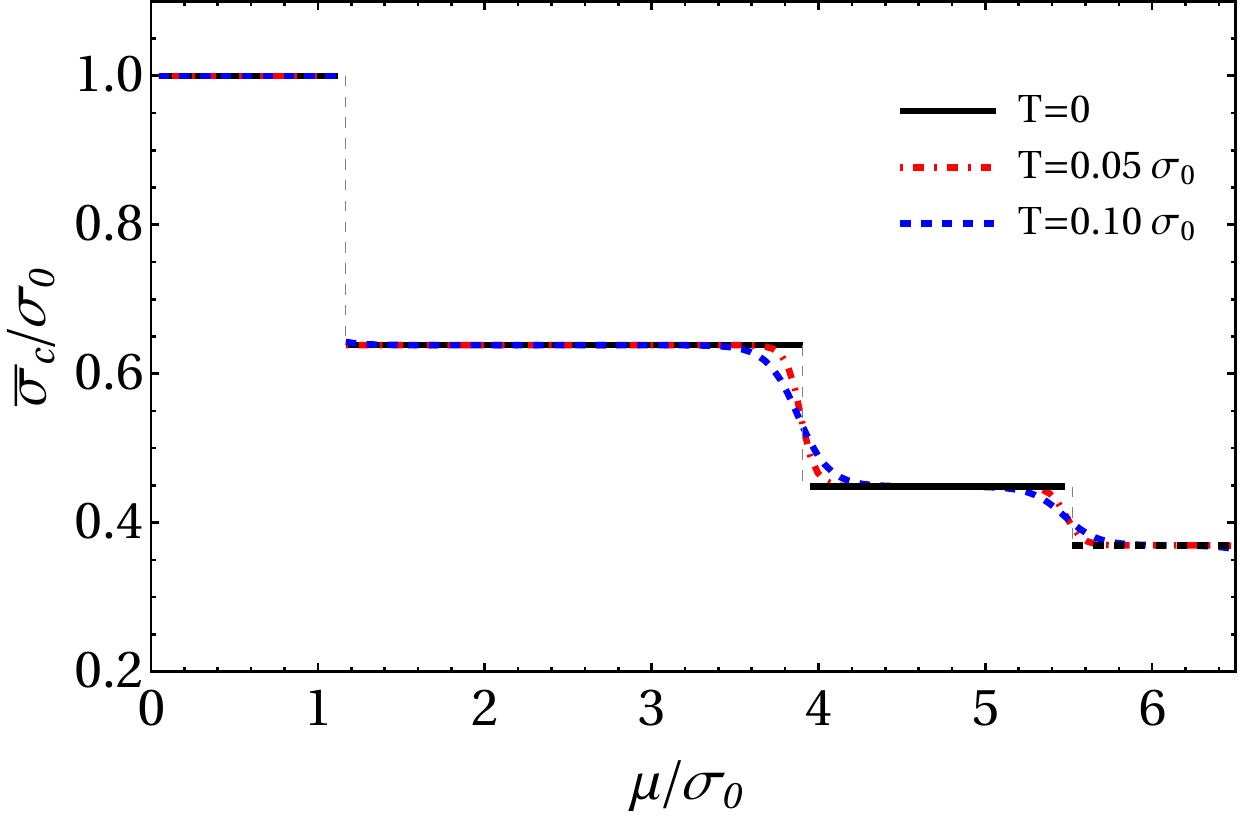}}
\subfigure[]{\includegraphics[width=7.5cm]{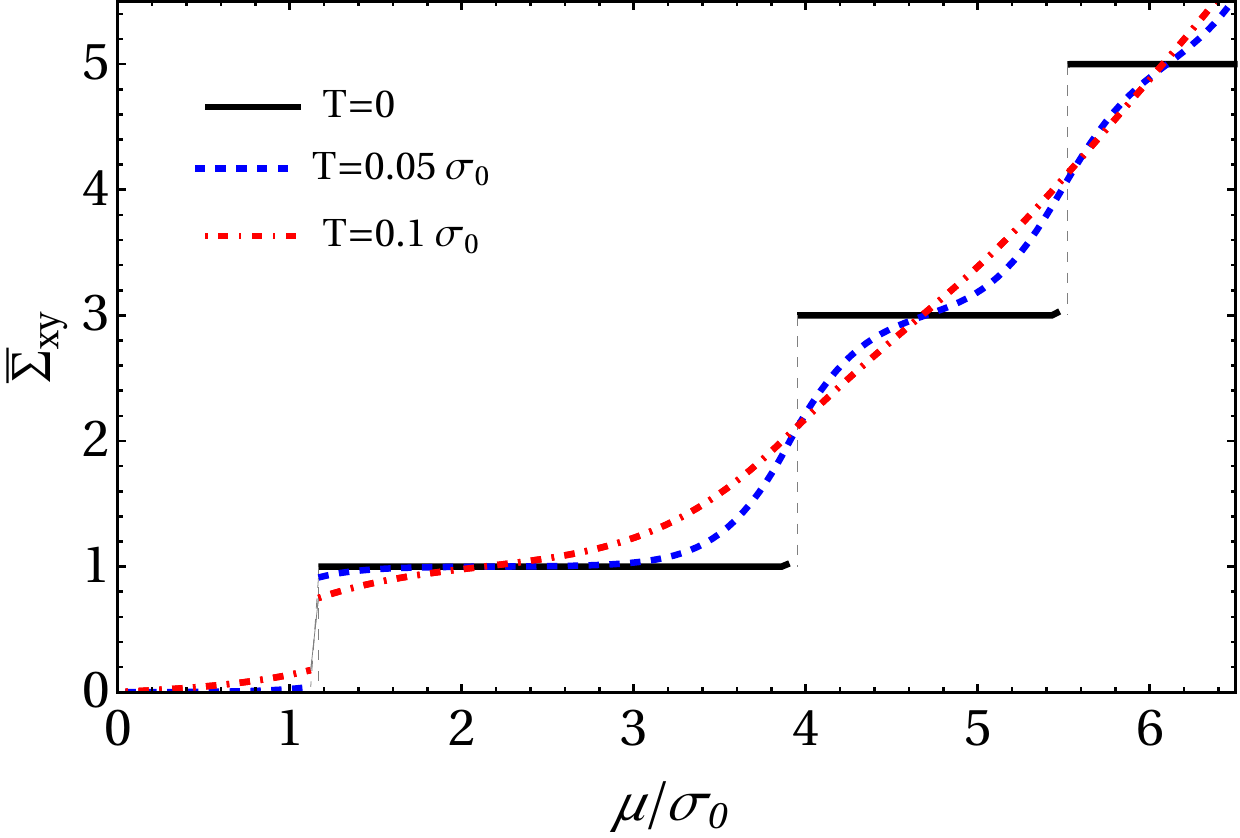}}
 \caption{Results for (a) the normalized EI condensate,
 (b) the scalar condensate, and (c) for the Hall conductivity, all as functions of chemical potential for several temperatures at fixed
 $eB=7.5\sigma_0^2$.}
 \label{fig12}
\end{figure}
\end{center}

{}For the parameter set considered here, the onset of the Hall response
coincides with the first-order emergence of the EI condensate. This
correlation is made explicit in {}Fig.~\ref{fig11}. Within the EI phase,
$\bar\eta_c$ is finite and the total gap remains larger than $\mu$, so
that there is no positive-energy Landau level and
$\bar\Sigma_{xy}=0$. At $\mu_c$, the EI condensate collapses, the total
gap decreases discontinuously, and the lowest Landau level becomes
occupied. The Hall conductivity therefore jumps to its first nonzero
plateau. Magnetic catalysis is visible both in the higher value of
$\bar\eta_c$ and in the field-dependent displacement of this onset.
{}For the parameter values considered here, the Hall conductivity becomes 
finite at the same chemical potential at which the EI condensate disappears. 
This coincidence results from the self-consistently determined quasiparticle spectrum 
and is not a general requirement. More generally, the onset of a finite Hall response 
is governed by the Landau-level occupation condition given in Eq.~(\ref{Hallthresholdresults}).

The effects of finite temperature are illustrated in {}Fig.~\ref{fig12} for
$eB=7.5\sigma_0^2$. Over the small temperature range shown, the EI
condensate and its first-order transition chemical potential change
only weakly. The behavior after the EI transition is more sensitive to
temperature. The discontinuities of the scalar condensate are rounded,
and the sharp Hall plateaus are replaced by smooth crossovers because
the Fermi--Dirac distribution thermally broadens the Landau-level
occupation thresholds. Once $\bar\eta_c=0$, the remaining gap that enters
the Hall response is controlled by $\bar\sigma_c$, which explains the
close correspondence between the middle and lower panels.

\begin{figure}[!htb]
 \centering
 \includegraphics[width=0.91\linewidth]{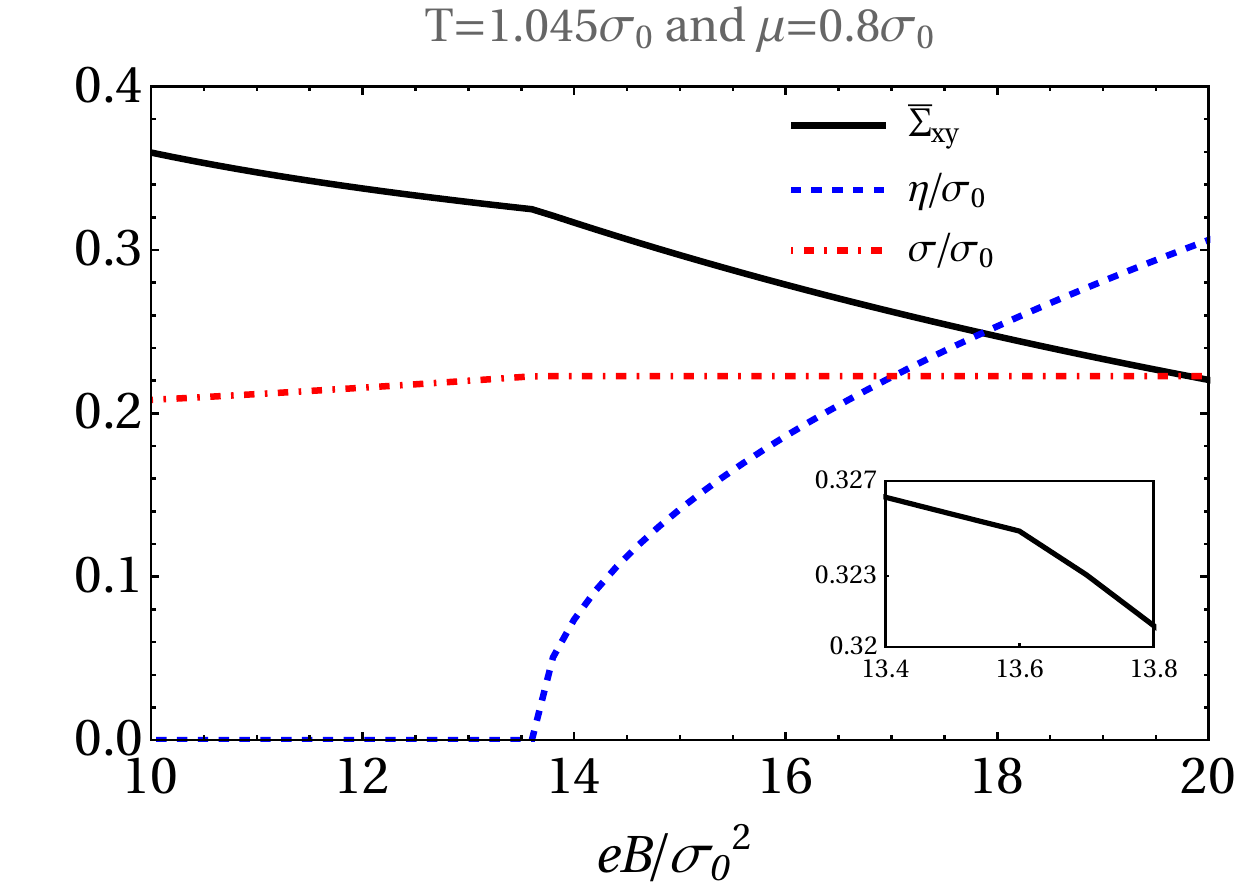}
 \caption{Normalized Hall conductivity (black solid
 curve), EI condensate (blue dashed curve), and scalar condensate (red
 dash-dotted curve) as functions of magnetic field at
 $T=1.045\sigma_0$ and $\mu=0.8\sigma_0$. At
 $eB_c\simeq13.6\sigma_0^2$, the EI condensate emerges continuously,
 the scalar condensate becomes pinned at a value $\sigma \simeq 0.22 \sigma_0$, and the Hall
 conductivity exhibits a change of slope. The inset enlarges the Hall
 response near the critical field.}
 \label{fig13}
\end{figure}

{}Finally, {}Fig.~\ref{fig13} examines the field-driven onset of EI order
at fixed $T=1.045\sigma_0$ and $\mu=0.8\sigma_0$. These values are chosen in such a way as to be close to the transition shown in {}Fig.~\ref{fig9} when $eB \gtrsim 10 \sigma_0^2$. {}For
$eB<eB_c\simeq13.6\sigma_0^2$, the equilibrium solution has
$\bar\eta_c=0$. In this regime, the scalar condensate increases through
magnetic catalysis, whereas the Hall conductivity decreases with the
field. At $eB=eB_c$, the EI condensate emerges continuously. At the
same point, the scalar condensate becomes pinned at a value $\sigma \simeq 0.22 \sigma_0$,
while the Hall curve develops a visible slope change. {}For
$eB>eB_c$, the EI condensate grows with the field and the Hall
conductivity continues to decrease approximately linearly over the
range shown.

The near-critical behavior is analyzed in the Appendix~\ref{AppC}. Within
the present mean-field approximation,
\begin{align}
 \bar\eta_c
 &\propto
 (eB-eB_c)^{1/2},
 \label{etaCriticalScaling}\\
 \left|
 \bar\Sigma_{xy}-\bar\Sigma_{xy,c}
 \right|
 &\propto
 eB-eB_c,
 \label{HallCriticalScaling}
\end{align}
for $eB\gtrsim eB_c$. The EI exponent $1/2$ is the usual mean-field
order-parameter exponent. The linear Hall scaling follows because the
leading change in the fermionic spectrum and the Hall response is
analytic in $\bar\eta_c^{\,2}$ near the continuous transition. The
slope change in {}Fig.~\ref{fig13} is therefore a transport signature of
the onset of the EI order within this model, complementary to the direct
observation of the excitation gap.

\subsection{Comparison with experimental condensed-matter materials}
\label{sec:experimental_comparison}

The above results have been expressed in terms of the intrinsic energy
scale $\sigma_0$ and the dimensionless magnetic field $eB/\sigma_0^2$.
A comparison with condensed-matter systems requires restoring both
$\hbar$ and the effective Dirac velocity $v_F$. {}For an isotropic Dirac cone,
the dimensionless field used in the numerical analysis is related to
the magnetic field in SI units by
\begin{equation}
 \frac{eB}{\sigma_0^2}
 \;\longrightarrow\;
 \frac{\hbar |e| B_{\rm SI}v_F^2}{\sigma_0^2}.
 \label{fieldconversionmaterials}
\end{equation}
For an anisotropic spectrum, $v_F^2$ in
Eq.~(\ref{fieldconversionmaterials}) is replaced by $v_xv_y$, so that
$v_F^{\rm eff}=\sqrt{v_xv_y}$. Numerically (expressing the magnetic field 
in units of tesla),
\begin{equation}
 B_{\rm SI}[{\rm T}]
 \simeq
 0.1519
 \left(\frac{eB}{\sigma_0^2}\right)
 \left(\frac{\sigma_0}{1\ {\rm meV}}\right)^2
 \left(\frac{10^5\ {\rm m\,s^{-1}}}{v_F^{\rm eff}}\right)^2 .
 \label{fieldconversiontesla}
\end{equation}
This expression is equivalent to using
$1~{\rm T}\simeq692.4~{\rm eV}^2$ and
$e\simeq1/\sqrt{137}$, with the factor $v_F/c$ restored consistently.
The conversion of temperature is made with
$k_B=8.6173\times10^{-5}\ {\rm eV\,K^{-1}}$.

The identification of $\sigma_0$ with an experimental energy scale
also requires some care. For the parameter set used in most of the
figures, the zero-field solution gives
$\bar\eta_c(0)\simeq0.8\sigma_0$ and
$T_c(0)\simeq0.67\sigma_0$. The corresponding full quasiparticle gap
at zero momentum is therefore
\begin{equation}
 \Delta_{\rm qp}(0)
 =2\sqrt{\sigma_0^2+\bar\eta_c^{\,2}(0)}
 \simeq2.56\sigma_0.
 \label{modelgapmaterials}
\end{equation}
Consequently, the comparisons below should be understood as
representative calibrations of the model rather than as quantitative
fits to a specific sample.

A particularly favorable case is provided by InAs/GaSb quantum wells.
Transport and terahertz spectroscopy measurements reported an
excitonic gap of approximately $2~{\rm meV}$ and a bulk critical
temperature of approximately $10~{\rm K}$~\cite{Du}. Taking
$\sigma_0\simeq1~{\rm meV}$ gives
$\Delta_{\rm qp}(0)\simeq2.56~{\rm meV}$ and
$T_c(0)\simeq7.8~{\rm K}$, both reasonably close to the measured
scales. {}For the velocity, we use the representative Bernevig--Hughes--Zhang parameter $A=30.5~{\rm meV\,nm}$ obtained from a band-structure
model of an InAs/GaSb quantum-well structure~\cite{BostromInAs}, which
corresponds to
\begin{equation}
 v_F=\frac{A}{\hbar}\simeq4.6\times10^4~{\rm m\,s^{-1}}.
\end{equation}
At $T=10~{\rm K}$ one has $T/\sigma_0\simeq0.862$. Reading the
corresponding point from Fig.~\ref{fig7} gives
$eB_c/\sigma_0^2\simeq4.95$, and hence
\begin{equation}
 B_c^{\rm InAs/GaSb}(10~{\rm K})\simeq3.5~{\rm T}.
 \label{BcInAs}
\end{equation}
The change in slope in the Hall conductivity shown in
{}Fig.~\ref{fig13} occurs at
$eB_H/\sigma_0^2\simeq13.6$. {}For the same calibration, this gives
\begin{equation}
 B_H^{\rm InAs/GaSb}\simeq9.6~{\rm T}.
 \label{BHInAs}
\end{equation}
The parameter point of {}Fig.~\ref{fig13} corresponds in this case to
$T\simeq12.1~{\rm K}$ and $\mu\simeq0.8~{\rm meV}$. These field and
temperature scales are experimentally accessible, making InAs/GaSb the
most favorable of the representative systems considered here.

\begin{table*}[t]
 \centering
 \caption{Representative conversion of the dimensionless magnetic
 field to physical units. The values should be viewed as
 order-of-magnitude estimates because both $\sigma_0$ and the effective
 Dirac velocities are material- and sample-dependent.}
 \label{tab:materialfieldscales}
 \begin{ruledtabular}
 \begin{tabular}{lcccc}
 Material
 & $\sigma_0$ $({\rm meV})$
 & $v_F^{\rm eff}$ $({\rm m\,s^{-1}})$
 & $B_c$ $({\rm T})$
 & $B_H$ $({\rm T})$\\
 \hline
 InAs/GaSb
 & $1$
 & $4.6\times10^4$
 & $3.5$ at $10~{\rm K}$
 & $9.6$ \\
 monolayer WTe$_2$
 & $15$
 & $1.9\times10^5$
 & $49$ at $150~{\rm K}$
 & $1.3\times10^2$
 \end{tabular}
 \end{ruledtabular}
\end{table*}

Monolayer WTe$_2$ provides a second, genuinely two-dimensional
comparison. Measurements of the chemical potential found a step that
saturates at approximately $40~{\rm meV}$ at low temperature, while a
more strongly insulating state develops below approximately
$100~{\rm K}$; equilibrium excitons were inferred to persist to still
higher temperatures~\cite{Sun}. Choosing $\sigma_0\simeq15~{\rm meV}$
gives $\Delta_{\rm qp}(0)\simeq38~{\rm meV}$ and
$T_c(0)\simeq117~{\rm K}$, providing a consistent order-of-magnitude
calibration. A representative low-energy model of monolayer WTe$_2$
uses the anisotropic velocities
$\hbar v_x\simeq0.5~{\rm eV\,\mathring{A}}$ and
$\hbar v_y\simeq3~{\rm eV\,\mathring{A}}$~\cite{KwanWTe2}, which give
\begin{equation}
 v_F^{\rm eff}=\sqrt{v_xv_y}
 \simeq1.9\times10^5~{\rm m\,s^{-1}}.
\end{equation}
For this calibration, $100~{\rm K}$ corresponds to
$T/\sigma_0\simeq0.575<T_c(0)/\sigma_0$, so that the model is already
in the EI phase at zero magnetic field. As an illustrative nonzero field
benchmark, raising the critical temperature to $150~{\rm K}$ gives
$T/\sigma_0\simeq0.862$ and therefore, from Fig.~\ref{fig7},
\begin{equation}
 B_c^{\rm WTe_2}(150~{\rm K})\simeq49~{\rm T}.
 \label{BcWTe2}
\end{equation}
The Hall-slope feature of Fig.~\ref{fig13} would instead occur at
\begin{equation}
 B_H^{\rm WTe_2}\simeq1.3\times10^2~{\rm T},
 \label{BHWTe2}
\end{equation}
with the corresponding temperature and chemical potential being
approximately $182~{\rm K}$ and $12~{\rm meV}$, respectively. Such a
field is not accessible in conventional steady-field experiments, but
it lies in the range of specialized ultrahigh pulsed-field techniques.
The comparison with WTe$_2$ is therefore less favorable than with InAs/GaSb, but
it remains useful as an indication of the scales required to observe
field-induced changes in an atomically thin EI candidate.
The resulting estimates are summarized in
Table~\ref{tab:materialfieldscales}.

{}For comparison, Ta$_2$NiSe$_5$ has a much larger characteristic gap
and a transition near $326~{\rm K}$~\cite{Lu}. Since the magnetic
field inferred from Eq.~(\ref{fieldconversiontesla}) scales as
$\sigma_0^2/(v_F^{\rm eff})^2$, a direct application of the present
calibration leads to fields far larger than those obtained for the two
systems above. Moreover, its strongly anisotropic, quasi-one-dimensional
band structure and the simultaneous structural transition make the
mapping to the present planar Dirac model less controlled. Therefore, we
do not pursue a detailed numerical comparison for Ta$_2$NiSe$_5$.

These estimates are intended to identify the physical scale of the
magnetic effects predicted by the model. A more quantitative comparison
would require a material-specific determination of the effective
velocities, the relation between the measured spectral gap and
$\sigma_0$, and the inclusion of effects omitted here, such as Zeeman
splitting, disorder, finite thickness, and deviations from a strictly
Dirac-like dispersion. The inclusion of these details is beyond the scope
of the present paper.

\section{Conclusions}
\label{conclusions}

In this work, we investigated the effects of a constant magnetic field
perpendicular to the plane on the excitonic-insulator (EI) phase of an
extended Gross--Neveu model in $(2+1)$-dimensions. The model contains
a conventional (chiral) scalar interaction channel and an additional channel
associated with spontaneous interband coherence. Working at leading order in
the large-$N$ expansion, we derived the renormalized thermodynamic
potential at finite temperature, chemical potential, and magnetic
field and solved the coupled gap equations by comparing all stationary
solutions of the potential.

Our results show that the perpendicular magnetic field enhances the EI
condensate and enlarges the region in which the excitonic phase is
thermodynamically favored. At zero chemical potential, both the
zero-temperature EI gap and the critical temperature increase
monotonically with the field. The scalar condensate behaves
differently. It remains fixed at $\bar\sigma_c=\sigma_0$ throughout
the EI phase and acquires a nontrivial dependence on temperature,
chemical potential, and magnetic field only after the EI condensate
has vanished.

At zero temperature, the density-driven EI transition is first order.
Its critical chemical potential depends nonmonotonically on the
magnetic field. Weak-field oscillations and the initial decrease in
$\mu_c(B)$ arise from changes in Landau-level occupation, whereas the
subsequent increase at stronger fields reflects the dominance of
magnetic catalysis. The complete
$T$--$\mu$ phase diagrams show that increasing the field expands the EI
region and moves the mean-field tricritical point to higher
temperatures, thereby increasing the portion of the phase boundary
corresponding to a first-order transition.

We also analyzed the Hall response generated by the Landau-level
spectrum. At zero temperature, the Hall conductivity remains zero
while the chemical potential lies below the lowest positive-energy
Landau level and develops the expected sequence of plateaus as
successive levels become occupied. For the parameters considered here,
the first nonzero Hall plateau appears at the same first-order
transition at which the EI condensate disappears. This coincidence is
not a general identity, since the onset of a Hall response is governed
more generally by the Landau-level occupation threshold. At finite temperature, the plateaus and the
Landau-level-induced discontinuities are smoothed by thermal
occupation.

A complementary signature is obtained when the EI phase is induced by
the magnetic field at fixed temperature and chemical potential.  The emergence of the
EI condensate is accompanied by a visible change in the slope of the
Hall conductivity. Close to the transition, the large-$N$ mean-field
solution gives
$\bar\eta_c\propto(eB-eB_c)^{1/2}$, while the leading variation of the
Hall conductivity is linear in $eB-eB_c$. Thus, within the present
model, Hall measurements can provide information complementary to the
direct observation of a spectral gap and may help identify the onset
or disappearance of excitonic order.

We have also performed an order-of-magnitude comparison with experimental condensed-matter
systems. This comparison indicates that low-gap semiconductor heterostructures provide
the most favorable setting for observing the predicted magnetic
effects. 
Our numerical estimates for the condensed materials considered should be interpreted as representative
calibrations rather than material-specific fits. A quantitative
application would require a more detailed treatment of the microscopic
band structure and of the relation between the measured excitation gap
and the effective scale $\sigma_0$. 

It would also be important to
include effects not considered here, such as Zeeman splitting,
disorder, finite layer thickness, anisotropic or non-Dirac dispersion,
and the nonlocal electron-hole Coulomb interaction. 
In strictly
two-dimensional systems, collective phase fluctuations and possible
Berezinskii--Kosterlitz--Thouless physics may further separate the
pair-formation scale obtained in the homogeneous large-$N$
approximation from the actual phase-coherence temperature. 
We furthermore note that the model studied here can be also modified to treat different types of excitons, following for instance the generalized Gross-Neveu model considered in~\cite{Klimenko2016}, where a phase structure of a more general (2+1)-dimensional Gross--Neveu model with four coupling constants was investigated and that includes two bispinor structures, $i\bar\psi \gamma^5\psi$ and $i\bar\psi \gamma^3\psi$, which can act as EI order parameters. Extending
the present framework in these directions would allow a more direct
connection between magnetic catalysis, Hall transport, and excitonic
ordering in specific planar materials.

\section*{Acknowledgments}
W.R.T. gratefully acknowledges the kind hospitality of the Centro de F\'{\i}sica at the Universidade de Coimbra, 
where part of this work was conducted.
This work was partially supported by Conselho Nacional de Desenvolvimento Cient\'{\i}fico e Tecnol\'ogico (CNPq), 
grants No. 319307/2025-5 (R.O.R.) and 200037/2026-9 (W.R.T.); Funda\c{c}\~ao Carlos Chagas Filho de Amparo \`a 
Pesquisa do Estado do Rio de Janeiro (FAPERJ), Grants No. SEI-260003/019544/2022
(W.R.T.) and E-26/200.415/2026 (R.O.R.).

\appendix

\section{Renormalization of the effective potential}
\label{appA}

In this appendix, we derive the renormalized thermodynamic potential
used in Secs.~\ref{GN_model_2+1_d} and
\ref{GN_model_2+1_d_mag_field}. We first regularize the zero-field
fermion determinant and define cutoff-independent couplings. We then
introduce the shifted scalar field employed in Sec.~\ref{GN_model_2+1_d_mag_field}
and derive the finite source associated with the explicit semiconductor
gap. The ultraviolet divergence is independent of temperature,
chemical potential, and magnetic field. Consequently, the
renormalization prescription obtained at $T=\mu=B=0$ is sufficient for
all of the thermodynamic potentials considered in the main text.

Throughout the first part of this appendix, $\sigma_c$ denotes the
expectation value of the unshifted Hubbard--Stratonovich field appearing
in Sec.~\ref{GN_model_2+1_d}. The fermionic quasiparticle energy at
vanishing magnetic field is
\begin{equation}
 E_{\mathbf p}
 =
 \sqrt{\mathbf p^2+\rho_0^2},
 \qquad
 \rho_0^2
 =
 (m+\sigma_c)^2+\eta_c^2 .
 \label{appA:zeroBdispersion}
\end{equation}
The thermodynamic potential renormalization
is
\begin{align}
 \frac{\Omega(\sigma_c,\eta_c;T,\mu)}{N}
 &=
 \frac{\sigma_c^2}{2g_c}
 +
 \frac{\eta_c^2}{2g_e}
 \nonumber\\
 &-
 2T
 \sum_{\nu=-\infty}^{\infty}
 \int^{\Lambda}
 \frac{d^2p}{(2\pi)^2}
 \ln\!\left[
 (\omega_\nu-i\mu)^2+E_{\mathbf p}^2
 \right],
 \label{appA:unrenormalizedpotential}
\end{align}
where $\omega_\nu=(2\nu+1)\pi T$ and a sharp spatial-momentum cutoff
$\Lambda$ has been introduced. This is the regularized form of the
fermionic determinant written in Eq.~(\ref{Veff001}).
Performing the Matsubara sum gives, up to a field-independent additive
constant,
\begin{align}
 \frac{\Omega(\sigma_c,\eta_c;T,\mu)}{N}
 &=
 \frac{\sigma_c^2}{2g_c}
 +
 \frac{\eta_c^2}{2g_e}
 -
 2\int^{\Lambda}\frac{d^2p}{(2\pi)^2}E_{\mathbf p}
 \nonumber\\
 &-
 2T\int\frac{d^2p}{(2\pi)^2}
 \Bigg[
 \ln\!\left(1+e^{-(E_{\mathbf p}-\mu)/T}\right)
 \nonumber\\
 &
 +
 \ln\left(1+e^{-(E_{\mathbf p}+\mu)/T}\right)
 \Bigg].
 \label{appA:afterMatsubara}
\end{align}
The thermal terms are ultraviolet finite. The only divergence is
contained in the vacuum integral that we denote here by
\begin{align}
 I_{\rm vac}(\rho_0)
 &\equiv
 -2\int^{\Lambda}\frac{d^2p}{(2\pi)^2}
 \sqrt{\mathbf p^2+\rho_0^2}
 \nonumber\\
 &=-\frac{1}{\pi}
 \int_0^{\Lambda}dp\,p\sqrt{p^2+\rho_0^2}
 \nonumber\\
 &=-\frac{1}{3\pi}
 \left[
 (\Lambda^2+\rho_0^2)^{3/2}-\rho_0^3
 \right].
 \label{appA:vacuumexact}
\end{align}
{}For $\Lambda\gg\rho_0$, this becomes
\begin{equation}
 I_{\rm vac}(\rho_0)
 =
 -\frac{\Lambda^3}{3\pi}
 -\frac{\Lambda\rho_0^2}{2\pi}
 +\frac{\rho_0^3}{3\pi}
 +{\cal O}(\Lambda^{-1}).
 \label{appA:vacuumexpansion}
\end{equation}
The first term is independent of the condensates and can be removed by
an irrelevant normalization of the thermodynamic potential. Defining
\begin{equation}
 g_{\Lambda}\equiv\frac{\pi}{\Lambda},
 \qquad
 \frac{1}{g_{\Lambda}}=\frac{\Lambda}{\pi},
 \label{appA:gLambda}
\end{equation}
the regularized zero-temperature potential can be written as
\begin{align}
\frac{ \Omega_{\rm reg}(\sigma_c,\eta_c)}{N}
 &=
 \frac{\sigma_c^2}{2g_c}
 -
 \frac{(m+\sigma_c)^2}{2g_{\Lambda}}
 \nonumber\\
 &+
 \frac{\eta_c^2}{2}
 \left(
 \frac{1}{g_e}-\frac{1}{g_{\Lambda}}
 \right)
 +
 \frac{\left[(m+\sigma_c)^2+\eta_c^2\right]^{3/2}}{3\pi}.
 \label{appA:regularizedunshifted}
\end{align}
Equation~(\ref{appA:regularizedunshifted}) is the vacuum part of the
zero-field result displayed in Eq.~(\ref{VeffTmu}).

It is useful to absorb the explicit mass into the scalar gap by
introducing
\begin{equation}
\tilde{m}_c\equiv m+\sigma_c .
 \label{appA:shift}
\end{equation}
In terms of $\tilde{m}_c$, Eq.~(\ref{appA:regularizedunshifted}) becomes
\begin{align}
\frac{ \Omega_{\rm reg}(\tilde{m}_c,\eta_c)}{N}
 &=
 \frac{\tilde{m}^2_c}{2}
 \left(
 \frac{1}{g_c}-\frac{1}{g_{\Lambda}}
 \right)
 -
 \frac{m}{g_c}\tilde{m}_c
 \nonumber\\
 &+
 \frac{\eta_c^2}{2}
 \left(
 \frac{1}{g_e}-\frac{1}{g_{\Lambda}}
 \right)
 +
 \frac{(\tilde{m}^2_c+\eta_c^2)^{3/2}}{3\pi},
 \label{appA:regularizedshifted}
\end{align}
where the field-independent term $m^2/(2g_c)$ has been omitted. This
shift is the origin of Eq.~(\ref{sigmashift}). In
Sec.~\ref{GN_model_2+1_d_mag_field}, the shifted field $\Sigma_c$ is
relabelled as $\sigma_c$; hence, from that section onward,
$\sigma_c$ denotes the full scalar contribution to the quasiparticle
gap rather than the unshifted auxiliary field.

The cutoff dependence is removed by defining the renormalized,
scale-independent couplings
\begin{equation}
 \frac{1}{g_c^R}
 \equiv
 \frac{1}{g_c}-\frac{1}{g_{\Lambda}},
 \qquad
 \frac{1}{g_e^R}
 \equiv
 \frac{1}{g_e}-\frac{1}{g_{\Lambda}}.
 \label{appA:renormalizedcouplings}
\end{equation}
These relations are equivalent to imposing curvature normalization
conditions on the effective potential at arbitrary reference values of
the two background fields and subsequently subtracting the finite
reference-scale contributions. Since the main text is written directly
in terms of the scale-independent parameters $g_c^R$ and $g_e^R$, the
intermediate reference scales are not needed explicitly.

The combination multiplying the shifted scalar field is also finite.
We denote it by
\begin{equation}
 {\cal H}_c\equiv\frac{m}{g_c}.
 \label{appA:HcBare}
\end{equation}
The quantity ${\cal H}_c$ is the external source associated with the
explicit semiconductor gap. Thus, while the bare parameters $m$ and
$g_c$ separately depend on the cutoff, their ratio in
Eq.~(\ref{appA:HcBare}) is held fixed in the renormalized theory.

To express this source in terms of a physical vacuum scale, we define
$M$ as the stationary value of the shifted scalar field at
$T=\mu=B=0$ on the branch with $\eta_c=0$. The vacuum stationarity
condition is
\begin{equation}
 \left.
 \frac{\partial\Omega_{\rm reg}}
 {\partial\tilde{m}_c}
 \right|_{\tilde{m}_c=M,\,\eta_c=0}
 =0,
 \label{appA:Mcondition}
\end{equation}
which gives
\begin{equation}
 {\cal H}_c
 =
 \frac{m}{g_c}
 =
 \frac{M}{g_c^R}
 +
 \frac{M|M|}{\pi}.
 \label{A19}
\end{equation}
Equation~(\ref{A19}) is the mass-renormalization condition used in
Eqs.~(\ref{Hcdefinition}) and (\ref{Hcmassrelation}). It replaces the
cutoff-dependent bare mass by the finite reference scale $M$ and the
renormalized scalar coupling.

The fully renormalized zero-temperature and zero-density potential is
therefore
\begin{align}
\frac{ \Omega_{\rm R}(\tilde{m}_c,\eta_c)}{N}
 &=
 \frac{\tilde{m}_c^2}{2g_c^R}
 -
 {\cal H}_c\tilde{m}_c
 +
 \frac{\eta_c^2}{2g_e^R}
 \nonumber\\
 &+
 \frac{(\tilde{m}_c^2+\eta_c^2)^{3/2}}{3\pi}.
 \label{appA:renormalizedvacuum}
\end{align}
After the relabelling $\tilde{m}_c\to\sigma_c$ adopted in
Sec.~\ref{GN_model_2+1_d_mag_field}, Eq.~(\ref{appA:renormalizedvacuum})
is precisely the zero-field limit shown in Eq.~(\ref{veffT0}). At
finite temperature and chemical potential, the corresponding
zero-field result is
\begin{align}
 \frac{ \Omega_{\rm R}(\sigma_c,\eta_c;T,\mu)}{N}
 &=
 \frac{\sigma_c^2}{2g_c^R}
 -
 {\cal H}_c\sigma_c
 +
 \frac{\eta_c^2}{2g_e^R}
 +
 \frac{\rho^3}{3\pi}
 \nonumber\\
 &-
 \frac{T}{\pi}
 \int_0^\infty dp\,p
 \Bigg[
 \ln\!\left(1+e^{-(E_p-\mu)/T}\right)
 \nonumber\\
 & + \ln\!\left(1+e^{-(E_p+\mu)/T}\right)
 \Bigg],
 \label{appA:renormalizedfiniteT}
\end{align}
where
\begin{equation}
 \rho^2=\sigma_c^2+\eta_c^2,
 \qquad
 E_p=\sqrt{p^2+\rho^2}.
\end{equation}
This form makes the connection between the zero-field formulation in
Sec.~\ref{GN_model_2+1_d} and the magnetic-field potential in
Eq.~(\ref{VeffHTmu}) explicit.

Finally, the ultraviolet part of the fermion determinant is insensitive
to $T$, $\mu$, and the infrared Landau-level structure. Equivalently,
the difference between the magnetic and zero-field vacuum
contributions is finite. The counterterms encoded in
Eq.~(\ref{appA:renormalizedcouplings}) and the source condition
Eq.~(\ref{A19}) therefore also renormalize the theory at finite
magnetic field. This is why no additional $B$-dependent counterterms
appear in Sec.~\ref{GN_model_2+1_d_mag_field}.

\section{Scalar condensate in the excitonic-insulator phase}
\label{AppB}

In this appendix, we derive the relation used in
Sec.~\ref{numerical_results} to show that the scalar condensate remains
constant throughout the excitonic-insulator (EI) phase. Although the
result is most easily displayed using the finite-field gap equations,
it is independent of temperature, chemical potential, and magnetic
field within the homogeneous leading-order large-$N$ approximation.
We also give the corresponding zero-field vacuum value of the EI
condensate.

After the shift and relabelling described in
Sec.~\ref{GN_model_2+1_d_mag_field} and Appendix~\ref{appA}, the
renormalized thermodynamic potential per fermion species can be written
in the general form
\begin{equation}
 \frac{ \Omega_{\rm B}(\sigma_c,\eta_c;T,\mu,b)}{N}
 =
 \frac{\sigma_c^2}{2g_c^R}
 -{\cal H}_c\sigma_c
 +\frac{\eta_c^2}{2g_e^R}
 +{\cal F}(\rho;T,\mu,b),
 \label{AppB:generalpotential}
\end{equation}
where
\begin{equation}
 \rho^2=\sigma_c^2+\eta_c^2,
 \qquad
 {\cal H}_c
 =
 \frac{M}{g_c^R}
 +\frac{M|M|}{\pi}.
 \label{AppB:rhoH}
\end{equation}
The function ${\cal F}$ contains the vacuum, magnetic, and thermal
fermionic contributions displayed explicitly in
Eq.~(\ref{VeffHTmu}). The essential point is that these contributions
depend on the two condensates only through the invariant combination
$\rho$.

Defining
\begin{equation}
 {\cal K}(\rho;T,\mu,b)
 \equiv
 \frac{1}{\rho}
 \frac{\partial {\cal F}}{\partial\rho},
 \label{AppB:Kdefinition}
\end{equation}
the two stationarity conditions take the compact form
\begin{align}
 0
 &=
 \frac{\sigma_c}{g_c^R}
 -{\cal H}_c
 +\sigma_c{\cal K},
 \label{AppB:sigmagapcompact}\\
 0
 &=
 \eta_c
 \left(
 \frac{1}{g_e^R}+{\cal K}
 \right).
 \label{AppB:etagapcompact}
\end{align}
Equations~(\ref{AppB:sigmagapcompact}) and
(\ref{AppB:etagapcompact}) are equivalent to
Eqs.~(\ref{gap_equations_with_B_sigma}) and
(\ref{gap_equations_with_B_eta}), respectively. Writing the equations
in this form makes clear that the same function ${\cal K}$ appears in
both of them.

On the EI branch, $\eta_c\neq0$, and
Eq.~(\ref{AppB:etagapcompact}) implies
\begin{equation}
 {\cal K}=-\frac{1}{g_e^R}.
 \label{AppB:KEI}
\end{equation}
Substitution into Eq.~(\ref{AppB:sigmagapcompact}) then gives
\begin{equation}
 \sigma_c
 \left(
 \frac{1}{g_c^R}-\frac{1}{g_e^R}
 \right)
 =
 {\cal H}_c.
 \label{AppB:sigmarelation}
\end{equation}
Using Eq.~(\ref{AppB:rhoH}), one obtains
\begin{equation}
 \sigma_c
 =
 \frac{
 g_e^R M\left(g_c^R|M|+\pi\right)
 }{
 \pi\left(g_e^R-g_c^R\right)
 }
 \equiv\sigma_0 .
 \label{AppB:sigma0}
\end{equation}
This is Eq.~(\ref{sigmacEIconstant}) of the main text. Since all of the
$T$-, $\mu$-, and $b$-dependence is contained in the eliminated
function ${\cal K}$, Eq.~(\ref{AppB:sigma0}) remains valid throughout
the homogeneous EI phase. Hence, while $\eta_c\neq0$, the scalar
condensate is pinned to the constant value $\sigma_0$. The external
parameters modify the EI condensate and the total quasiparticle gap,
but not the scalar component itself. Once the EI solution disappears
and $\eta_c=0$, Eq.~(\ref{AppB:etagapcompact}) is satisfied identically,
the elimination leading to Eq.~(\ref{AppB:sigma0}) is no longer
possible, and $\sigma_c$ acquires the temperature-, density-, and
magnetic-field dependence shown in Sec.~\ref{numerical_results}.

{}For completeness, consider the EI solution at
$T=\mu=b=0$. In this limit, the fermionic part of the renormalized
potential is
\begin{equation}
 {\cal F}(\rho;0,0,0)=\frac{\rho^3}{3\pi},
\end{equation}
so that
\begin{equation}
 {\cal K}(\rho;0,0,0)=\frac{\rho}{\pi}.
 \label{AppB:Kvacuum}
\end{equation}
{}For a nonvanishing EI condensate,
Eqs.~(\ref{AppB:KEI}) and (\ref{AppB:Kvacuum}) yield
\begin{equation}
 \rho_0=-\frac{\pi}{g_e^R}.
 \label{AppB:rho0}
\end{equation}
The EI vacuum solution therefore requires $g_e^R<0$ and has
\begin{equation}
 \eta_0^2
 =
 \frac{\pi^2}{(g_e^R)^2}-\sigma_0^2,
 \qquad
 \eta_0
 =
 \pm\sqrt{\frac{\pi^2}{(g_e^R)^2}-\sigma_0^2}.
 \label{AppB:eta0}
\end{equation}
A real nonzero EI stationary solution consequently exists only when
$\pi^2/(g_e^R)^2>\sigma_0^2$. Thermodynamic stability still requires
that its value of the effective potential be compared with those of
the competing stationary branches, as done in the numerical analysis.
On the non-EI vacuum branch, $\eta_c=0$, the reference scale $M$ is the
positive scalar stationary solution by construction, through the
renormalization condition in Eq.~(\ref{A19}).

\section{Near-critical behavior of the Hall conductivity}
\label{AppC}

In this appendix, we analyze the behavior of the Hall conductivity in
the vicinity of the continuous magnetic-field-driven EI transition
shown in Fig.~\ref{fig13}. The purpose is to clarify why the onset of
the EI condensate produces a change in the slope of the Hall response
and to establish the corresponding mean-field scaling. We use the
notation introduced in Sec.~\ref{GN_model_2+1_d_mag_field},
\begin{equation}
 b\equiv |eB|,
 \qquad
 \alpha_n=2-\delta_{n0},
 \label{appC:definitions}
\end{equation}
and restrict the discussion to the positive fields used in
Fig.~\ref{fig13}, for which $b=eB$ and
$\operatorname{sgn}(eB)=1$.

At finite temperature, the normalized Hall conductivity defined in
Eq.~(\ref{normalizedHall}) is
\begin{equation}
 \bar\Sigma_{xy}(b)
 =
 \sum_{n=0}^{\infty}
 \alpha_n
 \left[
 f\!\left(\bar E_n(b)-\mu\right)
 -
 f\!\left(\bar E_n(b)+\mu\right)
 \right],
 \label{appC:HallFiniteT}
\end{equation}
where
\begin{equation}
 f(x)=\frac{1}{e^{x/T}+1},
\end{equation}
and the equilibrium Landau-level energies are
\begin{equation}
 \bar E_n(b)
 =
 \sqrt{
 2nb+\bar\sigma_c^{\,2}(b)+\bar\eta_c^{\,2}(b)
 }.
 \label{appC:LandauEnergy}
\end{equation}
Thus, the magnetic-field dependence of the Hall response is both
explicit, through the term $2nb$, and implicit, through the
self-consistent condensates.

Let $b_c$ denote the critical field of the continuous EI transition.
At the transition,
\begin{equation}
 \bar\eta_c(b_c)=0,
 \qquad
 \bar\sigma_c(b_c)=\sigma_0,
 \label{appC:criticalcondensates}
\end{equation}
for the parameter set considered in Fig.~\ref{fig13}. We define the
critical Landau-level energies by
\begin{equation}
 E_{n,c}
 \equiv
 \bar E_n(b_c)
 =
 \sqrt{2nb_c+\sigma_0^2},
 \label{appC:criticalenergies}
\end{equation}
and the Hall conductivity at the critical point by
\begin{equation}
\bar\Sigma_{xy,c}
 \equiv
 \bar\Sigma_{xy,c}(b_c)
 =
 \sum_{n=0}^{\infty}
 \alpha_n
 \left[
 f(E_{n,c}-\mu)-f(E_{n,c}+\mu)
 \right].
 \label{appC:criticalHall}
\end{equation}

We first approach the transition from the EI side, $b>b_c$. As shown
in Appendix~\ref{AppB}, the scalar condensate is pinned at
$\bar\sigma_c=\sigma_0$ whenever the EI condensate is nonzero. Writing
\begin{equation}
 \delta b\equiv b-b_c,
 \qquad
 \delta b>0,
 \label{appC:deltab}
\end{equation}
the Landau-level energies can be expanded as
\begin{align}
 \bar E_n(b)
 ={}&
 E_{n,c}
 +
 \frac{n}{E_{n,c}}\,\delta b
 +
 \frac{\bar\eta_c^{\,2}(b)}{2E_{n,c}}
 \nonumber\\
 &+
 {\cal O}\!\left(
 \delta b^2,
 \delta b\,\bar\eta_c^{\,2},
 \bar\eta_c^{\,4}
 \right).
 \label{appC:energyexpansion}
\end{align}
Expanding the Fermi--Dirac functions in
Eq.~(\ref{appC:HallFiniteT}) around $E_{n,c}\pm\mu$, we obtain that
\begin{align}
 \bar\Sigma_{xy}(b)-\bar\Sigma_{xy,c}
 ={}&
 D_c\,\delta b
 +
 C_c\,\bar\eta_c^{\,2}(b)
 \nonumber\\
 &+
 {\cal O}\!\left(
 \delta b^2,
 \delta b\,\bar\eta_c^{\,2},
 \bar\eta_c^{\,4}
 \right),
 \label{sigmabarminussigmac}
\end{align}
where
\begin{align}
 D_c
 \equiv{}&
 \sum_{n=0}^{\infty}
 \alpha_n
 \frac{n}{E_{n,c}}
 \left[
 f'(E_{n,c}-\mu)-f'(E_{n,c}+\mu)
 \right],
 \label{appC:Dc}\\
 C_c
 \equiv{}&
 \frac{1}{2}
 \sum_{n=0}^{\infty}
 \frac{\alpha_n}{E_{n,c}}
 \left[
 f'(E_{n,c}-\mu)-f'(E_{n,c}+\mu)
 \right],
 \label{appC:Cc}
\end{align}
and
\begin{equation}
 f'(x)=-\frac{1}{T}\frac{e^{x/T}}{\left(e^{x/T}+1\right)^2}.
\end{equation}
The term proportional to $D_c$ is the regular variation caused by the
explicit magnetic-field dependence of the Landau levels. The term
proportional to $C_c$ is the additional contribution generated by the
emergence of the EI gap. Keeping these two effects separate is
important: the variation of the Hall conductivity near $b_c$ is not
due solely to the EI condensate.

Within the mean-field large-$N$ approximation, the continuous EI
transition has the usual mean-field behavior
\begin{equation}
 \bar\eta_c(b)
 =
 A\,(b-b_c)^{\beta_{\rm EI}}
 +\cdots,
 \qquad
 \beta_{\rm EI}=\frac{1}{2},
 \label{appC:etaScaling}
\end{equation}
where $A$ is a nonuniversal amplitude. Substitution into
Eq.~(\ref{sigmabarminussigmac}) yields
\begin{equation}
 \bar\Sigma_{xy}(b)-\bar\Sigma_{xy,c}
 =
 \left(D_c+C_cA^2\right)(b-b_c)
 +{\cal O}\!\left[(b-b_c)^2\right].
 \label{criticalexp_Hall}
\end{equation}
Hence, provided that the coefficient $D_c+C_cA^2$ does not vanish
accidentally,
\begin{equation}
 \left|
 \bar\Sigma_{xy}(b)-\bar\Sigma_{xy,c}
 \right|
 \propto
 (b-b_c)^{\beta_H},
 \qquad
 \beta_H=1.
 \label{appC:HallExponent}
\end{equation}
Equivalently,
\begin{equation}
 \ln\left|
\bar\Sigma_{xy}-\bar\Sigma_{xy,c}
 \right|
 =
 \mathrm{const.}
 +
 \ln(b-b_c)
 +\cdots .
 \label{appC:logScaling}
\end{equation}
The linear behavior follows from the analyticity of the Hall response
in the squared EI order parameter together with
$\bar\eta_c^{\,2}\propto b-b_c$. It should therefore be understood as
a result of the present finite-temperature mean-field treatment rather
than as an independent universal critical exponent of the Hall
conductivity.

The change of slope displayed in Fig.~\ref{fig13} can be made more
explicit by also considering the normal side, $b<b_c$. There,
$\bar\eta_c=0$, while the scalar condensate varies with the field. If
\begin{equation}
 s_-
 \equiv
 \left.
 \frac{d\bar\sigma_c}{db}
 \right|_{b=b_c^-},
 \label{appC:sminus}
\end{equation}
then the one-sided slope below the transition is
\begin{equation}
 \left.
 \frac{d\bar\Sigma_{xy}}{db}
 \right|_{b=b_c^-}
 =
 D_c+S_c s_-,
 \label{appC:slopeBelow}
\end{equation}
with
\begin{equation}
 S_c
 \equiv
 \sum_{n=0}^{\infty}
 \alpha_n
 \frac{\sigma_0}{E_{n,c}}
 \left[
 f'(E_{n,c}-\mu)-f'(E_{n,c}+\mu)
 \right].
 \label{appC:Sc}
\end{equation}
Above the transition, Appendix~\ref{AppB} gives
$d\bar\sigma_c/db=0$, while the EI contribution is present, and hence
\begin{equation}
 \left.
 \frac{d\bar\Sigma_{xy}}{db}
 \right|_{b=b_c^+}
 =
 D_c+C_cA^2.
 \label{appC:slopeAbove}
\end{equation}
The difference between the one-sided slopes is consequently
\begin{equation}
 \left.
 \frac{d\bar\Sigma_{xy}}{db}
 \right|_{b=b_c^+}
 -
 \left.
 \frac{d\bar\Sigma_{xy}}{db}
 \right|_{b=b_c^-}
 =
 C_cA^2-S_c s_-.
 \label{appC:slopeChange}
\end{equation}
Thus, the slope change in Fig.~\ref{fig13} receives two related
self-consistent contributions: the onset of the EI condensate and the
simultaneous pinning of the scalar condensate at $\sigma_0$. Both
modify the quasiparticle spectrum entering the Hall response.

The expansion above assumes finite temperature, so that the
Fermi--Dirac functions are smooth. At strictly zero temperature, the
Hall conductivity is piecewise constant and changes discontinuously
when a Landau level crosses the chemical potential. In that limit, the
near-critical behavior must instead be analyzed using the occupation
thresholds in Eq.~(\ref{HallT0}).




\begin{thebibliography}{99}

\bibitem{Mott}
N. F. Mott,
The transition to the metallic state,
Philos. Mag. \textbf{6}, 287 (1961).

\bibitem{Knox}
R. S. Knox,
Theory of Excitons, Solid State Physics, Suppl. 5
(Academic Press, New York, 1963).

\bibitem{Des}
J. des Cloizeaux,
Excitonic instability and crystallographic anomalies in semiconductors,
J. Phys. Chem. Solids \textbf{26}, 259 (1965).

\bibitem{Keldysh}
L. V. Keldysh and Yu. V. Kopaev,
Possible instability of the semimetallic state toward Coulomb interaction,
Fiz. Tverd. Tela \textbf{6}, 2791 (1964)
[Sov. Phys. Solid State \textbf{6}, 2219 (1965)].

\bibitem{Jerome}
D. J\'erome, T. M. Rice, and W. Kohn,
Excitonic insulator,
Phys. Rev. \textbf{158}, 462 (1967).

\bibitem{Halperin}
B. I. Halperin and T. M. Rice,
Possible anomalies at a semimetal-semiconductor transition,
Rev. Mod. Phys. \textbf{40}, 755 (1968).

\bibitem{Zittartz}
J. Zittartz,
Theory of the excitonic insulator in the presence of normal impurities,
Phys. Rev. \textbf{162}, 752 (1967).

\bibitem{BCS}
J. Bardeen, L. N. Cooper, and J. R. Schrieffer,
Microscopic theory of superconductivity,
Phys. Rev. \textbf{106}, 162 (1957).

\bibitem{Bronold}
F. X. Bronold and H. Fehske,
Possibility of an excitonic insulator at the semiconductor-semimetal transition,
Phys. Rev. B \textbf{74}, 165107 (2006).

\bibitem{Comte}
C. Comte and P. Nozi\`eres,
Exciton Bose condensation: The ground state of an electron-hole gas. I. Mean-field description of a simplified model,
J. Phys. France \textbf{43}, 1069 (1982).

\bibitem{Du}
L. Du, {\it et al},
Evidence for a topological excitonic insulator in InAs/GaSb bilayers,
Nat. Commun. \textbf{8}, 1971 (2017).

\bibitem{Eisenstein}
J. P. Eisenstein and A. H. MacDonald,
Bose--Einstein condensation of excitons in bilayer electron systems,
Nature (London) \textbf{432}, 691 (2004).

\bibitem{Gupta}
S. Gupta, A. Kutana, and B. I. Yakobson,
Heterobilayers of two-dimensional materials as a platform for excitonic superfluidity,
Nat. Commun. \textbf{11}, 2989 (2020).

\bibitem{Ma}
L. Ma, {\it et al},
Strongly correlated excitonic insulator in atomic double layers,
Nature (London) \textbf{598}, 585 (2021).

\bibitem{Jia}
Y. Jia, P. Wang, C.-L. Chiu, \textit{et al.},
Evidence for a monolayer excitonic insulator,
Nat. Phys. \textbf{18}, 87 (2022).

\bibitem{Lu}
Y. F. Lu, {\it et al},
Zero-gap semiconductor to excitonic insulator transition in Ta$_2$NiSe$_5$,
Nat. Commun. \textbf{8}, 14408 (2017).

\bibitem{Volkov}
P. A. Volkov, {\it et al},
Critical charge fluctuations and emergent coherence in a strongly correlated excitonic insulator,
npj Quantum Mater. \textbf{6}, 52 (2021).

\bibitem{Sugimoto}
K. Sugimoto, S. Nishimoto, T. Kaneko, and Y. Ohta,
Strong-coupling nature of the excitonic insulator state in Ta$_2$NiSe$_5$,
Phys. Rev. Lett. \textbf{120}, 247602 (2018).

\bibitem{Kim}
K. Kim, H. Kim, J. Kim, C. Kwon, J. S. Kim, and B. J. Kim,
Direct observation of excitonic instability in Ta$_2$NiSe$_5$,
Nat. Commun. \textbf{12}, 1969 (2021).

\bibitem{Salvo}
F. J. Di Salvo, {\it et al},
Physical and structural properties of the new layered compounds Ta$_2$NiS$_5$ and Ta$_2$NiSe$_5$,
J. Less-Common Met. \textbf{116}, 51 (1986).

\bibitem{Wakisaka}
Y. Wakisaka, {\it et al},
Excitonic insulator state in Ta$_2$NiSe$_5$ probed by photoemission spectroscopy,
Phys. Rev. Lett. \textbf{103}, 026402 (2009).

\bibitem{Wakisaka2}
Y. Wakisaka, \textit{et al.},
Photoemission spectroscopy of Ta$_2$NiSe$_5$,
J. Supercond. Nov. Magn. \textbf{25}, 1231 (2012).

\bibitem{Windgatter}
L. Windg\"atter, {\it et al},
Common microscopic origin of the phase transitions in Ta$_2$NiS$_5$ and the excitonic insulator candidate Ta$_2$NiSe$_5$,
npj Comput. Mater. \textbf{7}, 210 (2021).

\bibitem{Varsano}
D. Varsano, M. Palummo, E. Molinari, and M. Rontani,
A monolayer transition-metal dichalcogenide as a topological excitonic insulator,
Nat. Nanotechnol. \textbf{15}, 367 (2020).

\bibitem{Sun}
B. Sun,  \textit{et al.},
Evidence for equilibrium exciton condensation in monolayer WTe$_2$,
Nat. Phys. \textbf{18}, 94 (2022).

\bibitem{Liu}
J. Liu, H. Qu, and Y. Li,
One-dimensional magnetic excitonic insulators,
New J. Phys. \textbf{26}, 103034 (2024).

\bibitem{Littlewood}
P. B. Littlewood and X. Zhu,
Possibilities for exciton condensation in semiconductor quantum-well structures,
Phys. Scr. \textbf{T68}, 56 (1996).

\bibitem{Lizardo}
L. H. C. M. Nunes, R. L. S. Farias, and E. C. Marino,
Superconducting and excitonic quantum phase transitions in doped Dirac electronic systems,
Phys. Lett. A \textbf{376}, 779 (2012).

\bibitem{Mazza}
G. Mazza,
Magnetic field control of the excitonic transition in Ta$_2$NiSe$_5$,
arXiv:2601.23136.

\bibitem{lopes}
N. Lopes, M. A. Continentino, and D. G. Barci,
Excitonic insulators and Gross--Neveu models,
Phys. Rev. B \textbf{105}, 165125 (2022).

\bibitem{Elliott}
R. J. Elliott and R. Loudon,
Theory of the absorption edge in semiconductors in a high magnetic field,
J. Phys. Chem. Solids \textbf{15}, 196 (1960).

\bibitem{fenton}
E. W. Fenton,
Excitonic insulator in a magnetic field,
Phys. Rev. \textbf{170}, 816 (1968).

\bibitem{Yafet}
Y. Yafet, R. W. Keyes, and E. N. Adams,
Hydrogen atom in a strong magnetic field,
J. Phys. Chem. Solids \textbf{1}, 137 (1956).

\bibitem{Zhu}
Z. Zhu, {\it et al},
Magnetic field tuning of an excitonic insulator between the weak- and strong-coupling regimes in quantum-limit graphite,
Sci. Rep. \textbf{7}, 1733 (2017).

\bibitem{mit1}
D. V. Khveshchenko,
Magnetic-field-induced insulating behavior in highly oriented pyrolitic graphite,
Phys. Rev. Lett. \textbf{87}, 206401 (2001).

\bibitem{mit2}
E. V. Gorbar, V. P. Gusynin, V. A. Miransky, and I. A. Shovkovy,
Magnetic-field-driven metal-insulator phase transition in planar systems,
Phys. Rev. B \textbf{66}, 045108 (2002).

\bibitem{mit3}
H. Leal and D. V. Khveshchenko,
Excitonic instability in two-dimensional degenerate semimetals,
Nucl. Phys. B \textbf{687}, 323 (2004).

\bibitem{hall1}
V. P. Gusynin and S. G. Sharapov,
Unconventional integer quantum Hall effect in graphene,
Phys. Rev. Lett. \textbf{95}, 146801 (2005).

\bibitem{hall2}
V. P. Gusynin, V. A. Miransky, S. G. Sharapov, and I. A. Shovkovy,
Excitonic gap, phase transition, and quantum Hall effect in graphene,
Phys. Rev. B \textbf{74}, 195429 (2006).

\bibitem{supercond1}
G. W. Semenoff, I. A. Shovkovy, and L. C. R. Wijewardhana,
Phase transition induced by a magnetic field,
Mod. Phys. Lett. A \textbf{13}, 1143 (1998).

\bibitem{supercond2}
K. Krishana, N. P. Ong, Y. Zhang, \textit{et al.},
Quasiparticle thermal Hall angle and magnetoconductance in YBa$_2$Cu$_3$O$_x$,
Phys. Rev. Lett. \textbf{82}, 5108 (1999).

\bibitem{supercond3}
D. V. Khveshchenko and W. F. Shively,
Excitonic pairing between nodal fermions,
Phys. Rev. B \textbf{73}, 115104 (2006).

\bibitem{Graphene}
K. S. Novoselov, {\it et al},
Two-dimensional gas of massless Dirac fermions in graphene,
Nature (London) \textbf{438}, 197 (2005).

\bibitem{gn}
D. J. Gross and A. Neveu,
Dynamical symmetry breaking in asymptotically free field theories,
Phys. Rev. D \textbf{10}, 3235 (1974).

\bibitem{Klimenko1}
K. G. Klimenko,
Three-dimensional Gross--Neveu model at nonzero temperature and in an external magnetic field,
Theor. Math. Phys. \textbf{90}, 1 (1992)
[Teor. Mat. Fiz. \textbf{90}, 3 (1992)].

\bibitem{Klimenko2}
K. G. Klimenko,
Three-dimensional Gross--Neveu model at nonzero temperature and in an external magnetic field,
Z. Phys. C \textbf{54}, 323 (1992).

\bibitem{Klimenko3}
K. G. Klimenko,
Three-dimensional Gross--Neveu model in an external magnetic field,
Theor. Math. Phys. \textbf{89}, 1161 (1992)
[Teor. Mat. Fiz. \textbf{89}, 211 (1991)].

\bibitem{Gusynin1}
V. P. Gusynin, V. A. Miransky, and I. A. Shovkovy,
Catalysis of dynamical flavor symmetry breaking by a magnetic field in (2+1) dimensions,
Phys. Rev. Lett. \textbf{73}, 3499 (1994)
[Erratum: Phys. Rev. Lett. \textbf{76}, 1005 (1996)].

\bibitem{Gusynin2}
V. P. Gusynin, V. A. Miransky, and I. A. Shovkovy,
Dynamical flavor symmetry breaking by a magnetic field in (2+1) dimensions,
Phys. Rev. D \textbf{52}, 4718 (1995).

\bibitem{Vshivtsev:1995ug}
A. S. Vshivtsev, B. V. Magnitsky, and K. G. Klimenko,
Spontaneous breaking of chiral invariance by an external magnetic field in (2+1) dimensions,
JETP Lett. \textbf{62}, 283 (1995).

\bibitem{Zhukovsky:2000yd}
V. C. Zhukovsky, K. G. Klimenko, and V. V. Khudyakov,
Magnetic catalysis in a $P$-even, chiral-invariant three-dimensional model with four-fermion interaction,
Theor. Math. Phys. \textbf{124}, 1132 (2000).

\bibitem{Gomes:2023vvu}
Y.~M.~P.~Gomes, E.~Martins, M.~B.~Pinto and R.~O.~Ramos,
First-order phase transitions within Weyl type of materials at low temperatures,
Phys. Rev. B \textbf{108}, no.8, 085107 (2023).

\bibitem{Caldas:2009zz}
H.~Caldas and R.~O.~Ramos,
Magnetization of planar four-fermion systems,
Phys. Rev. B \textbf{80}, 115428 (2009).

\bibitem{Ramos:2013aia}
R. O. Ramos and P. H. A. Manso,
Chiral phase transition in a planar four-Fermi model in a tilted magnetic field,
Phys. Rev. D \textbf{87}, 125014 (2013).

\bibitem{Klimenko:2013gua}
K. G. Klimenko and R. N. Zhokhov,
Magnetic catalysis effect in the (2+1)-dimensional Gross--Neveu model with Zeeman interaction,
Phys. Rev. D \textbf{88}, 105015 (2013).

\bibitem{Klimenko:2012qi}
K. G. Klimenko, R. N. Zhokhov, and V. C. Zhukovsky,
Superconductivity phenomenon induced by an external in-plane magnetic field in a (2+1)-dimensional Gross--Neveu-type model,
Mod. Phys. Lett. A \textbf{28}, 1350096 (2013).

\bibitem{Klimenko:2012tk}
K. G. Klimenko, R. N. Zhokhov, and V. C. Zhukovsky,
Superconducting phase transitions induced by chemical potential in a (2+1)-dimensional four-fermion quantum field theory,
Phys. Rev. D \textbf{86}, 105010 (2012).

\bibitem{Gomes:2022dmf}
Y.~M.~P.~Gomes and R.~O.~Ramos,
Superconducting phase transition in planar fermionic models with Dirac cone tilting,
Phys. Rev. B \textbf{107}, no.12, 125120 (2023).

\bibitem{Kneur:2013cva}
J. L. Kneur, M. B. Pinto, and R. O. Ramos,
Phase diagram of the magnetized planar Gross--Neveu model beyond the large-$N$ approximation,
Phys. Rev. D \textbf{88}, 045005 (2013).

\bibitem{Mauldin:2026lcj}
J.~L.~P.~Mauldin and D.~H.~Rischke,
The $(2+1)$-dimensional Gross-Neveu-Yukawa model at finite temperature, density, and magnetic field within the Functional Renormalization Group,
[arXiv:2608.02280 [hep-ph]].

\bibitem{Lenz:2023wvk}
J. J. Lenz, M. Mandl, and A. Wipf,
Magnetic catalysis in the (2+1)-dimensional Gross--Neveu model,
Phys. Rev. D \textbf{107}, 094505 (2023).

\bibitem{Lenz:2023gsq}
J.~J.~Lenz, M.~Mandl and A.~Wipf,
Magnetized (2+1)-dimensional Gross-Neveu model at finite density,
Phys. Rev. D \textbf{108}, no.7, 074508 (2023).

\bibitem{Shovkovy:2012zn}
I. A. Shovkovy, Magnetic Catalysis: A Review. In: D, Kharzeev, 
K. Landsteiner, A. Schmitt, Hu Yee, (eds) Strongly Interacting Matter 
in Magnetic Fields. Lecture Notes in Physics, vol 871, pp. 13-49 
(Springer, Berlin, Heidelberg, 2013).

\bibitem{Miransky:2015ava}
V. A. Miransky and I. A. Shovkovy,
Quantum field theory in a magnetic field: From quantum chromodynamics to graphene and Dirac semimetals,
Phys. Rep. \textbf{576}, 1 (2015).

\bibitem{Kneur:2006ht}
J. L. Kneur, M. B. Pinto, and R. O. Ramos,
Critical and tricritical points for the massless two-dimensional Gross--Neveu model beyond large $N$,
Phys. Rev. D \textbf{74}, 125020 (2006).

\bibitem{Barducci:1994cb}
A. Barducci, R. Casalbuoni, M. Modugno, G. Pettini, and R. Gatto,
Thermodynamics of the massive Gross--Neveu model,
Phys. Rev. D \textbf{51}, 3042 (1995).

\bibitem{Klimenko1988}
K. G. Klimenko,
Massive Gross--Neveu model in the leading order of the $1/N$ expansion: Allowance for temperature and chemical potential,
Theor. Math. Phys. \textbf{75}, 487 (1988).

\bibitem{njl1}
Y. Nambu and G. Jona-Lasinio,
Dynamical model of elementary particles based on an analogy with superconductivity. I,
Phys. Rev. \textbf{122}, 345 (1961);

\bibitem{njl2}
Y. Nambu and G. Jona-Lasinio,
Dynamical model of elementary particles based on an analogy with superconductivity. II,
Phys. Rev. \textbf{124}, 246 (1961).

\bibitem{chiralsym}
T. W. Appelquist, M. J. Bowick, D. Karabali, and L. C. R. Wijewardhana,
Spontaneous chiral-symmetry breaking in three-dimensional QED,
Phys. Rev. D \textbf{33}, 3704 (1986).

\bibitem{Flachi}
A. Flachi,
Strongly interacting fermions and phases of the Casimir effect,
Phys. Rev. Lett. \textbf{110}, 060401 (2013).

\bibitem{kittel}
C. Kittel,
\textit{Introduction to Solid State Physics}, 6th ed.
(Wiley, New York, 1986).

\bibitem{BostromInAs}
F. Vi\~nas Bostr\"om, A. Tsintzis, M. Hell, and M. Leijnse,
Band structure and end states in InAs/GaSb core-shell-shell nanowires,
Phys. Rev. B \textbf{102}, 195434 (2020).

\bibitem{KwanWTe2}
Y. H. Kwan, T. Devakul, S. L. Sondhi, and S. A. Parameswaran,
Theory of competing excitonic orders in insulating WTe$_2$ monolayers,
Phys. Rev. B \textbf{104}, 125133 (2021).

\bibitem{Klimenko2016}
D. Ebert, K. G. Klimenko, P. B.Kolmakov and V. C. Zhukovsky, Phase transitions in hexagonal, graphene-like lattice sheets and nanotubes under the influence of external conditions, Annals Phys. \textbf{371}, 254-286 (2016) doi:10.1016/j.aop.2016.05.001 [arXiv:1509.08093 [cond-mat.mes-hall]]

\end{thebibliography}

\end{document}